\documentclass[reprint, amsmath, amssymb, aps, superscriptaddress, pra]{revtex4-2}
\usepackage{format}
\allowdisplaybreaks

\begin{document}

\preprint{APS/123-QED}

\title{ Theory of itinerant collisional spin dynamics in nondegenerate molecular gases }

\author{Reuben R. W. Wang}
\email{reuben.wang@cfa.harvard.edu}
\affiliation{ ITAMP, Center for Astrophysics $|$ Harvard \& Smithsonian, Cambridge, Massachusetts 02138, USA }
\affiliation{ Department of Physics, Harvard University, Cambridge, Massachusetts 02138, USA }

\author{John L. Bohn}
\affiliation{JILA, NIST, and Department of Physics, University of Colorado, Boulder, Colorado 80309, USA}

\date{\today} 

\begin{abstract}

We study the fully itinerant dynamics of ultracold but nondegenerate polar molecules with a spin-$1/2$ degree of freedom encoded into two of their electric field dressed rotational states. 
Center of mass molecular motion is constrained to two-dimensions via tight confinement with a one-dimensional optical lattice, but remains mostly unconstrained within the plane. 
The pseudospins can become entangled through ultracold dipolar collisions, for which the locality of interactions is greatly relaxed by free molecular motion. 
At the level of single-molecule observables, collision-induced entanglement manifests as spin decoherence, for which our theoretical calculations serve well to describe recent Ramsey contrast measurements of quasi-2D confined KRb molecules at JILA [A. Carroll et al., Science 388 6745 (2025)]. 
In presenting a more detailed theoretical analysis of the KRb experiment, we highlight a key finding that molecular loss enhanced by particle exchange symmetry can lead to a suppression of collective spin decoherence, a mechanism with refer to as ``loss-induced quantum autoselection". 
We then show that by utilizing bialkali species with sufficiently large dipole moments, loss can be near completely suppressed in all collision channels via electric field tunable confinement-induced collisional shielding. 
The afforded collisional stability permits fully coherent spin mixing dynamics, natively realizing unitary circuit dynamics with random all-to-all connectivity and
U(1) charge conservation. 
This work establishes a bridge between the domains of ultracold molecular collisions and many-body spin physics, 
ultimately proposing the use of nondegenerate bulk molecular gases as a controllable platform for nonequilibrium explorations of itinerant quantum matter.   

\end{abstract}

\maketitle

\section{ \label{sec:introduction} Introduction }

Experimental advances in state preparation \cite{Picard24_PRXQ, Ruttley24_PRXQ} and optical confinement of large low entropy ensembles of ultracold molecules \cite{Yan13_Nat, DeMarco19_Sci, Valtolina20_Nat, Schindewolf22_Nat, Bigagli24_Nat}, provide a versatile platform for implementing and benchmarking theoretical models describing strongly interacting nonequilibrium quantum dynamics \cite{Buchler07_PRL, Carr09_NJP, Gorshkov11_PRL, Langen24_NatPhys, Cornish24_NatPhys}. 
On theoretical fronts, there has been tremendous progress in modeling and understanding the physics of lattice confined polar molecules \cite{Gorshkov11_PRA, Kruckenhauser20_PRA}, and classically controlled molecular interactions by movements in optical tweezer arrays \cite{Zhang22_PRXQ, Wang22_PRXQ, Sroczynska22_NJP}.   
In this paper, we address a regime less explored thus far, where internal molecular states (pseudo-spins) undergo coherent quantum dynamics, while the center-of-mass motion of these molecules (spin-carriers) undergo incoherent classical dynamics.
The former can influence the latter in expectation, whereas we argue that an adiabatic treatment applies to the converse in so much as to maintain separability of the spin and motional wavefunctions.  
This investigation draws motivation from recent experiments of two-dimensional (2D) layer confined KRb polar molecules \cite{Li23_Nat, Carroll25_Sci}, where the JILA group was able to observe itinerant spin dynamics with Ramsey spectroscopy measurements. 
Above, and for the remainder of this work, we will use itinerant to imply free center-of-mass molecular motion due to the absence of a corrugating lattice.

A central finding from our investigations is a mechanism we term \textit{loss-induced quantum autoselection}, proposed to explain a stretched exponential behavior of the Ramsey contrast decay observed in the JILA KRb experiment (JILA-KRb) \cite{Carroll25_Sci}. 
Contrary to the notion that particle loss invariably leads to decoherence, we show that loss processes can in fact dynamically steer the system into a population-depleted subspace that extends quantum coherence, extending the lifetime of the measured Ramsey contrast $C(t)$.
As will be further detailed in Sec.~\ref{sec:loss_autoselection}, particle exchange symmetry and entanglement underpins the autoselection process, revealing opportunities for quantum state selective control of bimolecular stereochemistry \cite{deMiranda11_NatPhys}.

Using the JILA-KRb study as a launchpad, a principal goal of this paper is to establish the utility of nondegenerate bulk molecular gases as a platforms for coherent many-body spin physics. 
Key to realizing this utility is the suppression of molecular losses, where without any fine tuning of intermolecular interactions, ultracold collisions of bialkali molecules generally result in short-range sticking dynamics \cite{Mayle12_PRA, Croft14_PRA, Christianen19_PRA, Gregory19_NatCom, Bause21_PRR, Bause23_JPCA} or chemical reaction \cite{Zuchowski10_PRA, Byrd10_PRA, Meyer10_PRA, Hu19_Sci, deMiranda11_NatPhys}, both of which lead to trap loss. 
By polarizing the molecules with a sufficiently large static electric field along $z$, tight coaxial optical confinement can induce collisional shielding by enforcing side-to-side repulsive dipole-dipole interactions that heavily suppresses molecular loss \cite{Micheli07_PRA, Quemener11_PRA, Zhu13_PRA, Valtolina20_Nat}. 
Such collisional stability not only reduces loss, but also protects molecular internal degrees of freedom from entangling with the external ones, that is essential to relaxing the temperature requirements for unitary spin dynamics.     
Moreover, itinerance of the molecules introduces an inherent degree of spatial and temporal noise, effectively realizing a class of random quantum circuit models \cite{Nahum17_PRX, Fisher23_ARCMP}. These models are shown in this paper to exhibit all-to-all connectivity, while conserving a global U(1) charge, connecting nondegenerate molecular gases to a broader landscape of theoretical models relevant to quantum information, entanglement and disordered dynamics.

The remainder of this paper is structured as instructions to building the aforementioned circuit from its components, sequentially addressing one, two, and many-body molecular processes with pedagogical care. 
In Sec.~\ref{sec:single_molecule}, we present the field-dressed rigid rotor model relevant to ultracold bialkali molecules and detail how a spin-1/2 is encoded in its spectrum.  
Sec.~\ref{sec:scattering_theory} is concerned with two-molecule interactions, building up a scattering formalism with its corresponding operators and observables. 
Details of JILA-KRb and of modeling its many-body dynamics are given in Sec.~\ref{sec:manybody_dynamics}, followed by our proposal for implementing unitary quantum circuit dynamics on nondegenerate molecular platforms.  
Finally, the conclusions and outlooks of this work are drawn in Sec.\ref{sec:conclusion}.

\section{ Single molecules and state preparation \label{sec:single_molecule} }

We are concerned with the dynamics of a dilute gas of ultracold molecules, confined in a cylindrically symmetric optical dipole trap (ODT) \cite{Grimm00_AAMOP}.
Although generally interacting, this section focuses on the molecules at the individual level, with considerations toward their internal and external DoF.

\subsection{ Internal structure }

\begin{figure}[ht]
    \centering
    \includegraphics[width=\columnwidth]{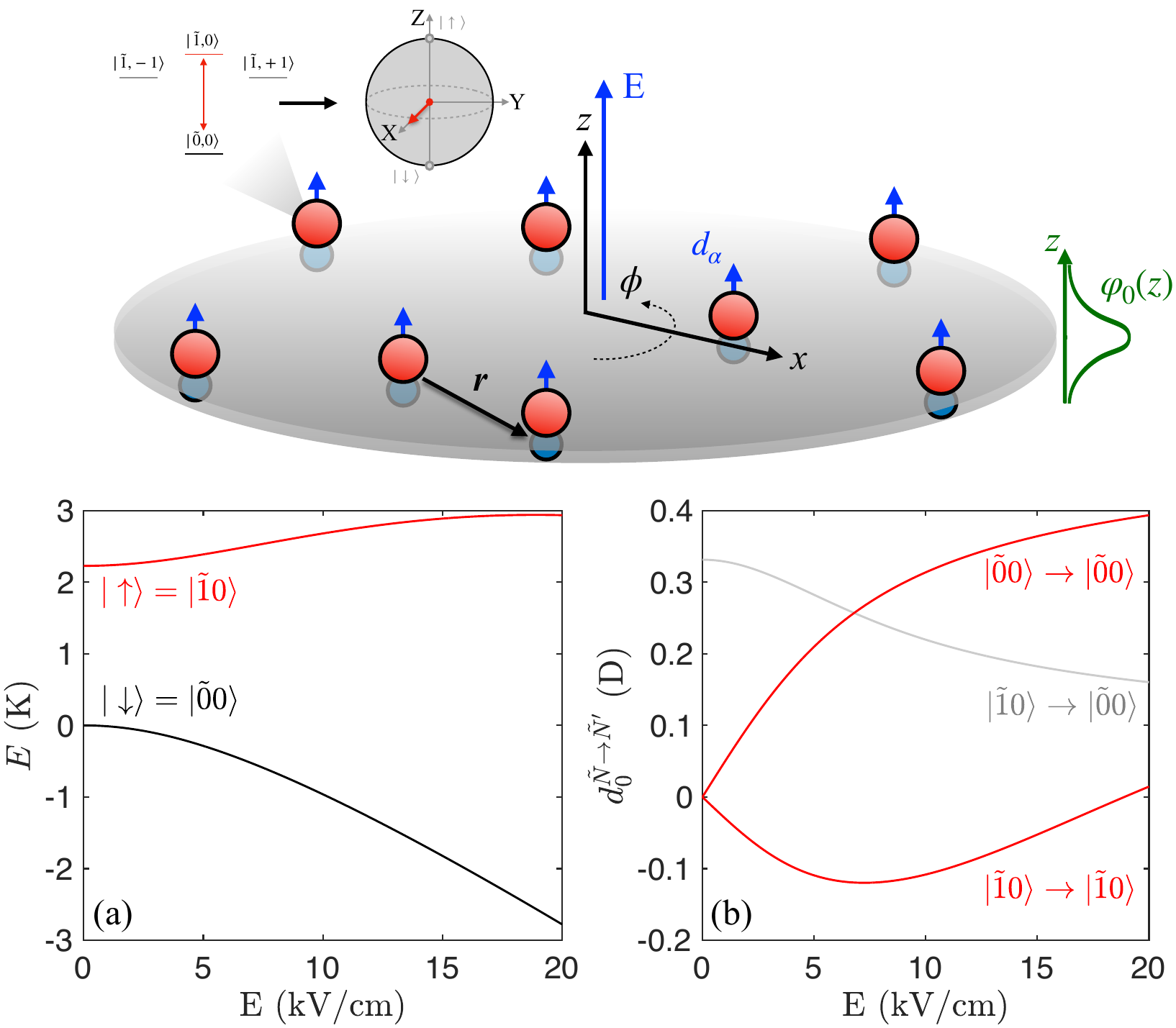}
    \caption{ (a) The dressed state energies of the $\ket{\uparrow}$ and $\ket{\downarrow}$ states as a function of electric field for KRb molecules under the rigid-rotor approximation. 
    (b) Induced (red) and transition (gray) dipole moments as a function of electric field for KRb molecules under the rigid-rotor approximation. The corresponding state-to-state transitions for each dipole moment are labeled in the figure. 
    Above the subplots is a summary illustration of the molecular gas platform, without consideration of their intermolecular interactions. 
    Tight optical confinement allows the molecules to remain in their harmonic oscillator groundstate $\varphi_0(z)$ along $z$. 
    The electric field ${\rm E}$ induces a state-dependent dipole moment $d_{\alpha}$, for the two lowest $M=0$ rotational states mapped onto a spin-1/2 system represented on the Bloch sphere. 
    }
    \label{fig:KRb_spinEnergiesPlot}
\end{figure}

The molecules are assumed spin polarized in their $^{1}\Sigma$ hyperfine groundstate by means of a sufficiently large applied magnetic field, so that only the rotational structure is relevant to the itinerant collisional dynamics. Subject also to a large electric field $\boldsymbol{{\rm E}}$ oriented along $z$, the molecule's internal Hamiltonian in its own center-of-mass frame is given by
\begin{align} \label{eq:single_molecule_Hamiltonian}
    h_i(t) 
    &=
    h_i^0
    +
    h_i^{\rm DD}(t) \nonumber\\
    &=
    B_{\rm rot} \boldsymbol{N}_i^2
    -
    \boldsymbol{d}_{i} \cdot \boldsymbol{{\rm E}}
    +
    h_i^{\rm DD}(t),
\end{align}
where $B_{\rm rot}$ is the rotational constant, $\boldsymbol{N}_i$ is the vector operator of molecular rotations, $\boldsymbol{d}_{i}$ is the molecular dipole operator and $h_i^{\rm DD}(t)$ inolves a time-dependent dynamical decoupling pulse sequence applied to the molecules that will be described below.  
We first focus on $h_i^0$, which have as eigenstates the field-dressed rotational states $| \tilde{N}_i, M_i \rangle$ that are linear combinations of the bare rotational states $| N_i, M_i \rangle$. 
The specific linear combinations can be obtained by diagonalizing the matrix
\begin{align}
    \langle N'_i, M'_i |
    h_i^0
    | N_i, M_i \rangle
    &=
    B_{\rm rot} N_i (N_i + 1) \delta_{N'_i, N_i} \delta_{M'_i, M_i} \\
    &\:\: -
    d_{0} {\rm E} ( -1 )^{M_i}
    \sqrt{ (2 N_i + 1) (2 N'_i + 1) } \nonumber\\
    &\quad \times 
    \begin{pmatrix}
        N'_i & 1 & N_i \\
        0 & 0 & 0
    \end{pmatrix}
    \begin{pmatrix}
        N'_i & 1 & N_i \\
        -M'_i & 0 & M_i
    \end{pmatrix}, \nonumber
\end{align}
which also provides us the dressed rotational spectrum $\epsilon_{\tilde{N}_i, M_i}$. 
Above, $d_0$ is the molecular frame dipole moment.
The induced and transition dipole moments between various field dressed states at ${\rm E}= {\rm E}^*$ are then given via the matrix elements:
\begin{align}
    d_{M_i}^{ \tilde{N}_i \rightarrow \tilde{N}'_i }({\rm E}^*)
    &=
    -\left.
    \frac{ \partial }{ \partial {\rm E} }
    \langle N'_i, M'_i | h_i^0({\rm E}) | N_i, M_i \rangle
    \right|_{{\rm E}^*}.
\end{align}
In this work, a sufficiently large applied field ${\rm E}$ allows us to assume that only the $M_i = 0$ states are relevant to the dynamics.
The lowest two induced dipole moments (red) and the transition dipole moment between them (gray) with $M_i = 0$ are plotted as a function of electric field in Fig.~\ref{fig:KRb_spinEnergiesPlot}\textcolor{blue}{a}.

We associate all inelastic processes to molecular loss from the trap, so that only the induced dipole moments are relevant. As such, we introduce the lighter notation $d_{M_i = 0}^{ \tilde{N}_i \rightarrow \tilde{N}_i }({\rm E}) = d_{ \tilde{N}_i }({\rm E})$. 
With an anharmonic molecular rotational spectrum, we can encode an effective pseudospin-1/2 DoF into the $\ket{ \downarrow } = | \tilde{0}, 0 \rangle$ and $\ket{\uparrow} = | \tilde{1}, 0 \rangle$ field-dressed rotational molecular states, with corresponding energies $\epsilon_{\downarrow}$ and $\epsilon_{\uparrow}$ (see Fig.~\ref{fig:KRb_spinEnergiesPlot}\textcolor{blue}{b}).
This encoding is illustrated at the top of Fig.~\ref{fig:KRb_spinEnergiesPlot}. 
Although only the $\{ \ket{ \tilde{0}, 0 }, \ket{ \tilde{1}, 0 } \}$ subspace is considered, we utilize rotational states up to $N = 5$ to ensure convergence of the computed dipole moments for states in this subspace. 
For the remainder of this paper, we will simply refer to this encoded DoF as spin instead of pseudospin.

Using a microwave pulse, the molecules are initially prepared in an equal positive superposition of the two spin states $\ket{ \chi(0) } = \ket{+} = { (\ket{\downarrow} + \ket{\uparrow}) / \sqrt{2} }$, which evolves in time via  
\begin{align} \label{eq:precessing_state}
    \ket{ \chi(t) }
    &=
    \frac{ \ket{\downarrow} + e^{ -i ( \epsilon_{\uparrow} - \epsilon_{\downarrow} ) t / \hbar } \ket{\uparrow} }{ \sqrt{2} }.
\end{align}  
However, the need to account for this dynamical phase can be eliminated by means of dynamical decoupling pulse sequences \cite{Viola98_PRA}, with details of the exact Hamiltonian time variation consolidated in $h_i^{\rm DD}(t)$. In essence, these pulses continually transform $\ket{\downarrow}$ into $\ket{\uparrow}$ and vice versa to serve as a feedforward control protocol to ``echo away" unwanted rotations of the Bloch vector due to experimental imperfections.  
For instance, the state in Eq.~(\ref{eq:precessing_state}) following a pulse would then be transformed into 
\begin{align}
    \ket{ \chi(t') }
    &=
    \frac{ \ket{\uparrow} + e^{ -i ( \epsilon_{\uparrow} - \epsilon_{\downarrow} ) t' / \hbar } \ket{\downarrow} }{ \sqrt{2} } \nonumber\\
    &=
    e^{ -i ( \epsilon_{\uparrow} - \epsilon_{\downarrow} ) t' / \hbar } 
    \left(
    \frac{ e^{ +i ( \epsilon_{\uparrow} - \epsilon_{\downarrow} ) t' / \hbar } \ket{\uparrow} + \ket{\downarrow} }{ \sqrt{2} }
    \right) \nonumber\\
    &\equiv
    \frac{ \ket{\downarrow} + e^{ + i ( \epsilon_{\uparrow} - \epsilon_{\downarrow} ) t' / \hbar } \ket{\uparrow} }{ \sqrt{2} },
\end{align}
effectively undoing the phase accumulated in Eq.~(\ref{eq:precessing_state}) if allowed to evolve for the same amount of time,
where the last equality arises from the global phase invariance of states in quantum mechanics and primes on $t$ indicate the post-pulse time. 
Repeated application of the pulses over time thus results in zero net azimuthal motion of the spins on the Bloch sphere.
Experiments typically utilize a more involved sequence of pulses that are so-called ``universal", protecting spins from all sorts of unwanted rotations. Such details are not the focus of this paper, so we refer the reader elsewhere for more information \cite{Viola98_PRA, Viola03_PRL, Lidar14_JWS}.

\subsection{ Molecular center-of-mass motion }

In the laboratory frame, the molecules of mass $m$ have their center-of-mass traverse paths dictated by the ODT potential energy surface.   
We will refer to the coordinates describing these paths interchangeably as molecular motion or external DoF. 
If sufficiently deep, the molecules remain mostly near the trap minima, allowing us to approximate the ODT as a harmonic potential $V_{\rm ext} = \frac{1}{2} m [ \omega_{\perp}^2 ( x^2 + y^2 ) + \omega_{z}^2 z^2]$, with radial and axial trap frequencies $\omega_{\perp}$ and $\omega_z$ respectively. 
The trap is tightly confining along $z$, such that the molecular vertical $z$ motion is energetically constrained to the groundstate harmonic oscillator state $\varphi_0(z)$ in the absence of interactions.   
This constraint places the gas in a quasi two-dimensional (quasi-2D) regime, which we assume for the remainder of this paper.

Albeit ultracold, we will assume that the kinetic energy of the molecules far surpasses $\hbar\omega_{\perp}$, such that the number of occupiable radial harmonic oscillator states greatly exceeds the actual number of molecules. 
In JILA-KRb, $\omega_{\perp} = 2\pi \times 39$ Hz which is three orders of magnitude smaller than $\omega_{z}= 2\pi \times 10$ kHz.  
Consequently, the molecules are mostly in far separated distinguishable motional states, so that quantum statistics plays a negligible role in their itinerant dynamics except during a collision. 
This point will be made precise in section \ref{sec:manybody_dynamics}. 
In such a nondegenerate gas, we expect that the 2D motional DoF $\boldsymbol{\xi}_i = \{ \boldsymbol{q}_i, \boldsymbol{p}_i \}$ for molecule $i$, to be described by a finite-temperature Gibbs state \cite{Polkovnikov10_AP}: 
\begin{align}
    \rho_{\boldsymbol{\xi}_i}(0)
    &=
    \frac{ 1 }{ Z }
    e^{-\beta H_i(\boldsymbol{q}_i, \boldsymbol{p}_i) } 
    \varphi_0(z),
\end{align}
where $Z$ is the partition function, $\beta = (k_B {\rm T})^{-1}$ is the inverse temperature and $H_i = \boldsymbol{p}_i^2/(2 m) + m \omega_{\perp}^2 \boldsymbol{q}_i^2/2$ is the Hamiltonian for the external DoF of molecule $i$. 
The initial total one-body density matrix is thus given by 
\begin{align} \label{eq:initial_density_matrix}
    \rho_i(0)
    &=
    \rho_{\boldsymbol{\xi}_i}(0)
    \otimes
    \rho_{\alpha_i}(0) \nonumber\\
    &=
    \frac{ 1 }{ Z }
    e^{-\beta H_i} 
    \varphi_0(z)
    \otimes 
    \ket{+}\bra{+}, 
\end{align}
where $\alpha$ labels the spin states $\ket{ \alpha } \in \{ \ket{\downarrow}, \ket{\uparrow} \}$. 
A summary illustration of the molecular setup discussed thus far is provided at the top of Fig.~\ref{fig:KRb_spinEnergiesPlot}.

\section{ Two-body interactions and scattering \label{sec:scattering_theory} }

With a large static electric field turned on, the long-range intermolecular interactions are strongly dipole-dipole in nature, but get overwhelmed by an attractive van der Waals interaction upon close approach: 
\begin{subequations}
\begin{align}
    V(\boldsymbol{R})
    &=
    -\frac{ C_6 }{ R^6 }
    +
    V_{\rm dd}(\boldsymbol{R}), \\
    V_{\rm dd}(\boldsymbol{R})
    &=
    \frac{ \boldsymbol{d}_1 \cdot \boldsymbol{d}_2 - 3 ( \boldsymbol{d}_1 \cdot \hat{\boldsymbol{R}} ) ( \boldsymbol{d}_2 \cdot \hat{\boldsymbol{R}} ) }{ 4 \pi \epsilon_0 R^3 },
\end{align}
\end{subequations}
where $\boldsymbol{R}$ is the three-dimensional relative coordinate between molecules $1$ and $2$, $C_6$ is the van der Waals coefficient \cite{Lepers13_PRA} and $\boldsymbol{d}_i$ is the dipole of molecule $i$.
Despite the long-range $1/R^3$ nature of $V_{\rm dd}$, the intermolecular interactions will be treated as effectively finite-ranged in this paper, by virtue of the nondegenerate temperatures that result in thermal energies $k_B {\rm T}$ far dominating the dipolar mean-field experienced per particle 
\footnote{ Quantitatively, we can estimate the dipolar mean-field interaction energy experienced per particle $\epsilon_{\rm mf}$ in a Maxwell-Boltzmann distributed gas with the analytic formula in Ref.~\cite{Lahaye09_IOP}.
We will consider the regime whereby $\epsilon_{\rm mf} \ll k_B {\rm T}$, which is the case for JILA-KRb with $\epsilon_{\rm mf} / (k_B {\rm T}) \lesssim 0.01$. 
In this regime, intermolecular interactions are only significant for two molecules that enter distances much shorter than the mean free path, treated here as scattering events.  
}.

The theory of ultracold molecular scattering is often handled with the time-independent Scr\"odinger equation, where two-body collisions are completely described by scattering phase shifts $\delta$, that modify the incident wavefunction while preserving unitarity.  
At low energies, Wigner's threshold law \cite{Sadeghpour00_JPB} tells us that three-dimensional scattering off a finite-range potential (i.e. $V \sim r^{-{\rm p}}$ with ${\rm p} > 3$) is dominated by $s$-waves, allowing a single parameter, the scattering length, to subsume all complications of the short range quantum dynamics.  
In much the same way, close-to-threshold dipolar scattering can be largely characterized by just the dipole length $a_{d, \tilde{N}} = m_r d_{ \tilde{N} }^2 / ( 4 \pi \epsilon_0 \hbar^2 )$.  
This simplification is especially true when the collision energy $E$, is far exceeded by the natural dipolar energy scale $E_{\rm dd} = \hbar^2 / ( m_r a_{d}^2 )$, dropping explicit reference to $\tilde{N}$ unless required for notational convenience. 

Since $E_{\rm dd} \propto d^{-4}$, molecules with larger dipole moments will have more stringent low temperature requirements for staying in the close-to-threshold regime \cite{Bohn09_NJP}.  
For KRb with a molecular frame dipole moment of $d = 0.574$ D \cite{Matsuda22_UCB}, even the largest electric fields of ${\rm E} \approx 12.7$ kV/cm considered here achieves a groundstate dipole moment of only $d_0 \approx 0.31$ D, corresponding to a dipolar energy of $E_{\rm dd} \approx 13$ kHz. 
With experimental temperatures of ${\rm T} \leq 300$ nK ($\approx 6$ kHz), we will assume the threshold scattering regime and utilize the relevant approximations therein.

\subsection{ Spatiotemporal translation invariance \label{sec:spatiotemporal_invariance} }

To ensure that our collisional formulation provides a consistent description of the itinerant spin dynamics, we will take special care to validate several assumptions normally presumed for bulk gases in the quantum collisional regime. 
In addition to bolstering agreement of our collisional model with observations in JILA-KRb, these details will be especially relevant when connecting itinerant molecular platforms to the circuit models discussed in Sec.~\ref{sec:unitary_dynamics}.  

A convenience normally afforded to ultracold scattering calculations is translational invariance in both space and time, so that the asymptotic scattering states are both collision momentum and energy eigenstates, labeled by the scattering channel index $\nu$. 
Scattering calculations then exploit these assumptions to derive closed-form analytic long-range scattering boundary conditions for matching of numerical close-coupling solutions \cite{Child74_AP, Quemener17_RSC}.  
This description is complicated if either of the translational invariances are broken, which would occur in the presence of external confinement and time-dependent driving.
Fortunate for the JILA-KRb, the full itinerant regime makes it possible to retain these invariances to a good approximation which we now justify. 

Bulk thermal molecular samples in optical dipole traps satisfy $k_B {\rm T} \gg \hbar\omega_{\perp}$, even at ultracold temperatures of $\sim 100$ nK. 
The external potential can thus be considered as slowly varying over the molecular thermal de Broglie wavelength, allowing us to make a Wentzel–Kramers–Brillouin (WKB) approximation to the eigenstates of $H_i$. The component of these eigenstates associated to the external DoF are then well approximated locally by momentum eigenstates $e^{i \boldsymbol{p}_{\rm local} \cdot \boldsymbol{r} / \hbar}$, with local momentum $\boldsymbol{p}_{\rm local} = \sqrt{ 2 m [E - V_{\rm ext}(r)] }$. So long as $\boldsymbol{p}_{\rm local}$ remains mostly constant over the collisional interaction region, we can treat completed collisions in the usual way by assuming a spatially translation invariant environment.    

In preserving quantum spin coherence, time translation invariance is also nominally broken by dynamical decoupling pulse sequences, commonplace to experimental spin platforms. 
In particular, JILA-KRb utilizes a symmetric Knill dynamical decoupling (KDD) sequence \cite{Souza11_PRL} that consists of 10 instantaneous $\pi$-pulses spaced by 50 $\mu$s for a total pulse block length of 0.5 ms. 
Collisions, on the other hand, also have their own intrinsic timescales related to the scattering phase shifts.  
In the quasi-2D regime with vertically polarized dipoles, the dipolar scattering phase shift is well approximated at threshold by the Born approximation as $\delta = -k a_d$ \cite{Ticknor09_PRA, Ticknor10_PRA}, allowing us an estimate of the Wigner collision time:
\begin{align}
    t_{\rm coll}
    &=
    \hbar
    \abs{ \frac{ d\delta }{ d E } }
    \approx 
    a_d (\hbar k / m_r)^{-1},
\end{align}
simply described as the free transit time across $a_d$, where $k = \sqrt{ 2 m_r E/ \hbar^2 }$ is the collision wavenumber with reduced mass $m_r = m / 2$. 
Scaling linearly with $a_d$, we find that the largest collision times probed by the KRb experiment are $\sim 10$ $\mu$s, while the smallest can be in the sub-nanosecond range.  
With collisions occurring much faster than the KDD pulse intervals, we will simply treat them as instantaneous so that they only occur ``in the dark" [i.e. subject to the time-independent Hamiltonian without $h_i^{\rm DD}(t)$].

Leveraging the WKB and instantaneous collision approximations above, we now proceed with a time-independent, momentum space formulation of scattering.

\subsection{ Scattering phase shifts of quasi-2D KRb \label{sec:scattering_phaseshifts} }

Threshold dipolar collisions that occur in quasi-2D have cross sections well approximated by the Born approximation \footnote{ Although usually applied to high energy collisions, the Born approximation is valid here based on the following physical argument: when close-to-threshold, the scatterers only see the long-range tail of the dipole-dipole potential before reflecting off the potential barrier. As a result, the molecules only experience weak dipole-dipole interactions in the effective interaction region, leaving their outgoing wavefunctions mostly unperturbed. 
Moreover, the Born approximation achieves the correct Wigner threshold laws as compared to full close-coupled scattering calculations. 
The Born approximation is valid in both two and three-dimensions for close-to-threshold dipolar collisions \cite{Ticknor10_PRA, Bohn14_PRA}. }. 
Of relevance to this work is a closely related scattering quantity, the scattering phase shift, which is accrued following a two-molecule encounter. 
Tabulated for the various collision channels into a single unitary operator, these phase shifts make up the scattering $S$-matrix $S$, that transforms a two-molecule density matrix from its pre-collision state $\rho$ into its post-collision state $\rho' = S \rho S^{\dagger}$ \cite{Newton79_FP}.           
The accrual of these channel-dependent phase shifts is what generates two-molecule entanglement \cite{Jaksch99_PRL, Brennen00_PRA}, consequently leading to single-molecule decoherence. 

With the dipolar molecules polarized along $z$, the two-body interactions are mostly repulsive at long-range where the adiabatic molecular channels remain spaced by $\sim$ GHz rotational energies, much larger than all other energy scales involved in the dynamics. 
Therefore, inelastic transitions between the various rotational channels at long-range, where most support of the wavefunction lies, are energetically suppressed. It is only when the molecules tunnel through the dipolar barrier into close proximity do significant exothermic inelastic processes or chemical reactions occur, both of which lead to rapid trap loss. 
Instead of treating these inelastic processes explicitly, we will simply impose an absorbing boundary condition behind the dipolar potential barrier. In this way, the $S$-matrix remains diagonal in the appropriately symmetrized basis:
\begin{subequations} \label{eq:2molecule_basis}
\begin{align}
    & \text{triplet sector}\:\:\left\{  
    \begin{tabular}{cc}
    $\ket{ \Downarrow }$
    &
    $= \ket{ \downarrow, \downarrow }$, \\[5pt]
    $\ket{ \Psi^{+} }$
    &
    $=\dfrac{ \ket{ \downarrow, \uparrow } + \ket{ \uparrow, \downarrow } }{ \sqrt{ 2 } }$, \\[10pt] 
    $\ket{ \Uparrow }$
    &
    $=\ket{ \uparrow, \uparrow }$
    \end{tabular} \right.
    \\
    & \text{singlet sector}\:\:\left\{
    \ket{ \Psi^{-} }
    =
    \frac{ \ket{ \downarrow, \uparrow } - \ket{ \uparrow, \downarrow } }{ \sqrt{ 2 } },
    \right.
\end{align}
\end{subequations}
but inherits complex valued phase shifts and loses unitarity. 
The suggestive notation of $\ket{ \Psi^{\pm} }$ is used to make clear that these are in fact maximally entangled Bell states. 
With molecules that are prepared in the $\ket{+}$ state, a first collision will only allow scattering in the triplet sector, since $\ket{+}^{\otimes 2} = ( \ket{ \Downarrow } + \sqrt{2} \ket{ \Psi^{+} } + \ket{ \Uparrow } ) / 2$. Subsequent collisions can, however, involve the singlet sector (see Sec.~\ref{sec:loss_autoselection}).

The $S$-matrix described above then treats each channel as effectively uncoupled during a collision,  
with the only relevant dipole matrix elements being the diagonal ones:  
\begin{subequations} \label{eq:DDI_scattering_terms}
\begin{align}
    \bra{ \Downarrow }
    V_{\rm dd}( \boldsymbol{r},z ) 
    \ket{ \Downarrow } 
    &=
    -2 C_{2,0}(\theta_{\rm E}) 
    \frac{ d_{\downarrow}^2 }{ 4 \pi \epsilon_0 }, 
    \\ 
    \bra{ \Psi^{+} }
    V_{\rm dd}( \boldsymbol{r},z ) 
    \ket{ \Psi^{+} } 
    &=
    -2 C_{2,0}(\theta_{\rm E})
    \frac{ ( d_{\downarrow} d_{\uparrow} + d_{\downarrow\uparrow}^2 ) }{ 4 \pi \epsilon_0 },
    \\ 
    \bra{ \Uparrow }
    V_{\rm dd}( \boldsymbol{r},z ) 
    \ket{ \Uparrow } 
    &=
    -2 C_{2,0}(\theta_{\rm E})
    \frac{ d_{\uparrow}^2 }{ 4 \pi \epsilon_0 }, \\
    \bra{ \Psi^{-} } 
    V_{\rm dd}( \boldsymbol{r},z ) 
    \ket{ \Psi^{-} }
    &=
    -2 C_{2,0}(\theta_{\rm E})
    \frac{ ( d_{\downarrow} d_{\uparrow} - d_{\downarrow\uparrow}^2 ) }{ 4 \pi \epsilon_0 },
\end{align}
\end{subequations}
where $C_{2,0}(\theta) = (3\cos^2\theta_{\rm E} - 1)/2$ is the reduced spherical harmonic, $\theta_{\rm E} = \cos^{-1}[ z (r^2 + z^2)^{-1/2} ]$ is the angle between $\boldsymbol{{\rm E}}$ and the relative two-molecule coordinate,  
and effective dipole moments were identified as
$d_{\downarrow} = \bra{ \downarrow } {d}_z \ket{ \downarrow }$, $d_{\uparrow} = \bra{ \uparrow } {d}_z \ket{ \uparrow }$ and $d_{\downarrow\uparrow} = \bra{ \downarrow } {d}_z \ket{ \uparrow } = \bra{ \uparrow } {d}_z \ket{ \downarrow }$. 
We also define the associated dipole lengths for each scattering channel:
\begin{subequations} \label{eq:effective_dipole_lengths}
\begin{align}
    a_{d, \Downarrow}
    &=
    \frac{ m_r d_{\Downarrow}^2 }{ 4 \pi \epsilon_0 \hbar^2 }
    =
    \frac{ m_r d_{\downarrow}^2 }{ 4 \pi \epsilon_0 \hbar^2 }, \\
    a_{d, {\Psi^{+}}}
    &=
    \frac{ m_r d_{+}^2 }{ 4 \pi \epsilon_0 \hbar^2 }
    =
    \frac{ m_r ( d_{\downarrow} d_{\uparrow} + d_{\downarrow\uparrow}^2 ) }{ 4 \pi \epsilon_0 \hbar^2 }, \\
    a_{d, \Uparrow}
    &=
    \frac{ m_r d_{\Uparrow}^2 }{ 4 \pi \epsilon_0 \hbar^2 }
    =
    \frac{ m_r d_{\uparrow}^2 }{ 4 \pi \epsilon_0 \hbar^2 }, \\
    a_{d, \Psi^{-}}
    &=
    \frac{ m_r d_{-}^2 }{ 4 \pi \epsilon_0 \hbar^2 }
    =
    \frac{ m_r ( d_{\downarrow} d_{\uparrow} - d_{\downarrow\uparrow}^2 ) }{ 4 \pi \epsilon_0 \hbar^2 }.
\end{align}
\end{subequations} 

Earlier in Sec.~\ref{sec:spatiotemporal_invariance}, an analysis of the Wigner collision time implied that scattering events can be treated as effectively instantaneous, occurring between KDD pulses.
However, the thermal states of molecular motion implies an uncertainty as to the instances at which collisions actually occur within a series of $\pi$-pulses. This uncertainty is quantified by the Heiseinberg inequality $\Delta E \Delta t \gtrsim \hbar$, where the energy uncertainty is set by the temperature of order $\sim$kHz. 
The time uncertainty is, therefore, bounded from below by $\Delta t \gtrsim 0.1$ ms, indicating an ambiguity of phases accrued on the $\ket{\Uparrow}$ or $\ket{\Downarrow}$ triplet state components in a molecular collision \footnote{
In Ref.~\cite{Miller24_Nat}, slower $100$ $\mu$s pulses were utilized for Floquet engineering of the interactions. An increase in spin decoherence at larger effective dipolar interactions was observed, compared to the case of native interactions achieved with static electric fields. We suspect this effect to have arisen discernible phase shifts between the $\ket{\Uparrow}$ and $\ket{\Downarrow}$ scattering channels. 
}.   
We thus introduce an effective time averaged dipole length: 
\begin{align} \label{eq:direct_dipole_length}
    a_{d, \updownarrow}
    &=
    \frac{ m_r d_{\updownarrow}^2 }{ 4 \pi \epsilon_0 \hbar^2 }
\end{align}
with $d_{\updownarrow}^2 = { (d_{\downarrow}^2 + d_{\uparrow}^2) / 2 }$,  
relevant to completed collisions in both $\ket{\Downarrow}$ and $\ket{\Uparrow}$ channels which we refer to as \textit{direct} interactions. 
Conversely, the scattering of anti-aligned spins will be referred to as \textit{exchange} interactions. 

\begin{figure}[ht]
    \centering
    \includegraphics[width=\columnwidth]{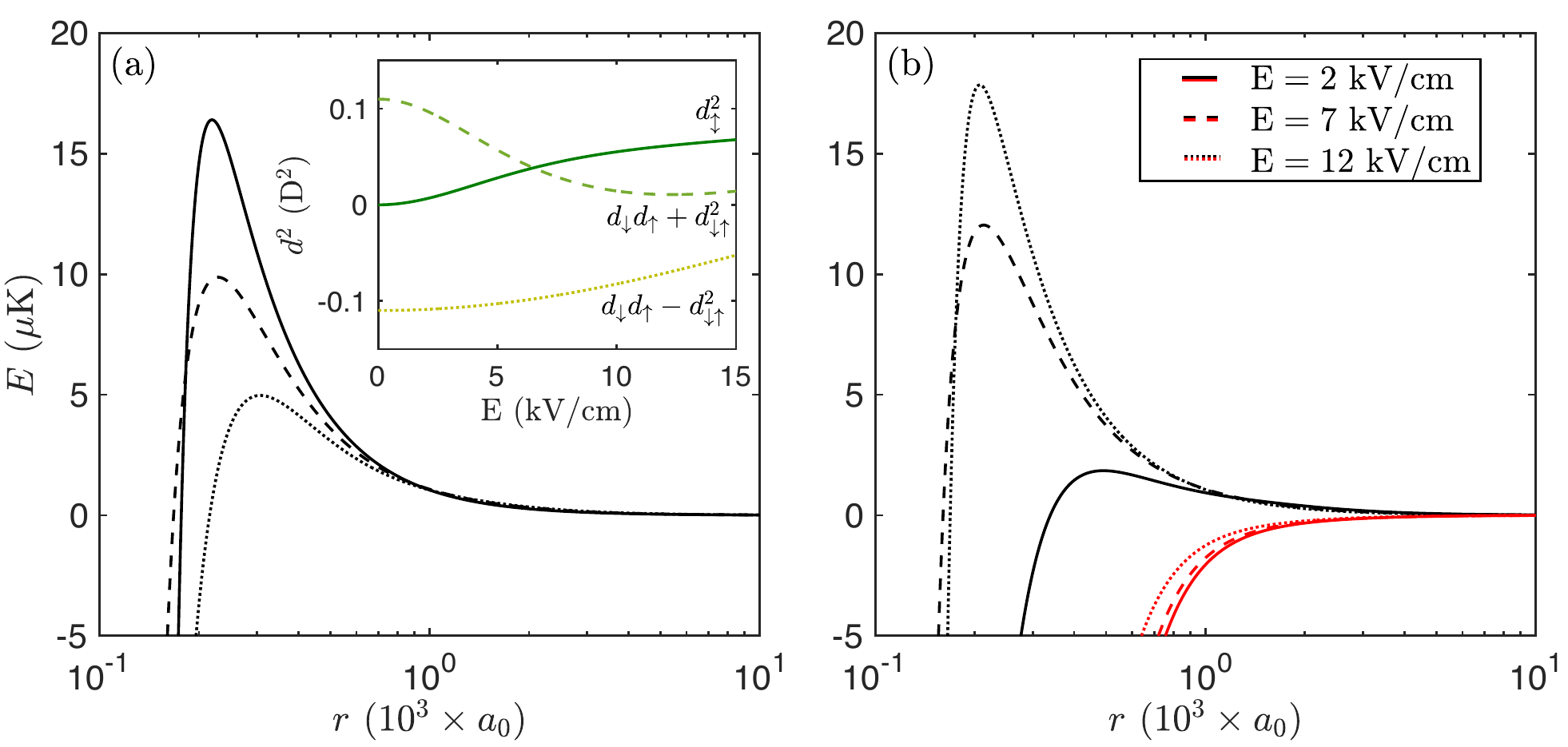}
    \caption{ The quasi-2D adiabatic curves in the (a) direct and (b) exchange scattering channels, for electric fields of ${\rm E} = 2$ kV/cm (solid), ${\rm E} = 7$ kV/cm (dashed) and $12$ kV/cm (dotted). 
    Subplot (b) plots the adiabats in both the triplet exchange channel with $p$-waves (black curves) and singlet exchange channels with $s$-waves (red curves).
    The effective direct (solid) and exchange (dashed and dotted) dipole moments are plotted as a function of ${\rm E}$ in the inset of subplot (a). 
        The $z$ trapping frequency is taken to be $\omega_z = 2 \pi \times 10$ kHz.  }
    \label{fig:KRb_potentials}
\end{figure}

The nature of two-molecule interactions is clearly depicted with plots of the adiabatic potential energy curves (adiabats).
These adiabats are obtained by first expanding the two-body Schr\"odinger equation in a cylindrical $\{ r, z, \phi \}$ separable basis $r^{-1/2} u(r) \ket{ n_z} \ket{ m_{\phi} }$, resulting in:
\begin{align} \label{eq:radial_SE}
    & 
    \frac{\hbar^2}{2 m_r} 
    \delta_{n'_z, n_z} 
    \delta_{m'_{\phi}, m_{\phi}}
    \left(
    \frac{ \partial^2 }{ \partial r^2 }
    -
    \frac{ m_{\phi}^2 - 1/4 }{ r^2 }
    +
    k_{\nu}^2
    \right)
    u(r) \\
    &=
    \bra{ n'_z, m'_{\phi} }
    \left(
    V_{\nu, \nu}
    +
    V_{{\rm ext}, z}
    -
    \frac{\hbar^2}{2 m_r}
    \frac{ \partial^2 }{ \partial z^2 }
    \right)
    \ket{ n_z, m_{\phi} }
    u(r), \nonumber
\end{align}
where $V_{{\rm ext}, z}(z) = V_{\rm ext}(\boldsymbol{0},z)$. 
The second derivative term in $r$ is then ignored, allowing us to diagonalize the remaining potential as a function of the in-plane radius $r$.
Further details of the derivation are given in App.~\ref{app:numerical_scattering}). 
Because $V_{\nu, \nu} + V_{{\rm ext}, z}$ is cylindrically symmetric, the $m_{\phi}$ partial wave quantum number is conserved so that no quantum coherence develops between internal angular momenta $|{ \tilde{N}_i, M_i }\rangle$ and external angular momenta $\ket{ m_{\phi} }$ during a collision. That is to say, the two molecule density matrix after a collision  has matrix elements that satisfy $\rho^{m'_{\phi}, m_{\phi}}_{ \nu', \nu } = 0$ when $m'_{\phi} \neq m_{\phi}$.
Consequently, taking a partial trace over $m_{\phi}$ is equivalent to an incoherent thermal average over external angular momenta, and is an operation that commutes with spin dynamics from elastic collisions \footnote{ Conservation of $m_{\phi}$ only remains strictly true if the wavefunction amplitude is truly negligible in the intermolecular short-range, otherwise the molecules could exert sufficient torques to reorient themselves and break the cylindrical symmetry. }. 
Following Fermi antisymmetry, we plot the adiabats with $p$-waves ($\abs{m_{\phi}} = 1$) for the symmetric triplet channels, and $s$-waves ($m_{\phi} = 0$) for the antisymmetric singlet channel in Fig.~\ref{fig:KRb_potentials}. 
At $\omega_z = 2\pi \times 10$ kHz used in JILA-KRb, the potential barrier height in triplet sector states does not monotonically increase with larger dipole moments, as was also seen in figure 3 of Ref.~\cite{Quemener11_PRA} due to an interplay of confinement-induced shielding and ``statistical suppression" between identical fermions.

For scattering in the triplet sector, we clearly see from Fig.~\ref{fig:KRb_potentials} that the interaction potentials are repulsive to ultracold molecules with the large $\gtrsim \mu$K barriers in the lowest $m_{\phi}=1$ partial wave. 
Therefore, molecules scattering in these channels primarily experience the long-range potential tails, which are well approximated by integrating out $z$ with just the ground harmonic oscillator state $\varphi_0(z)$.
Then along with the Born approximation, the diagonal $S$-matrix elements are given in real space as 
\begin{align} \label{eq:Smat_Born_approximation}
    {S}_{\nu, \nu}( \boldsymbol{k}', \boldsymbol{k} ) 
    &=
    e^{ 2 i \delta_{\nu}( \boldsymbol{k}', \boldsymbol{k} ) } \\
    &\approx 
    1
    -
    2 \pi i
    \bra{ \boldsymbol{k}' }
    \bra{ \nu }
    \int d z \abs{ \varphi_0(z) }^2
    {V}(\boldsymbol{r},z)
    \ket{ \nu }
    \ket{ \boldsymbol{k} }, \nonumber
\end{align}
with $\abs{ \boldsymbol{k}' } = \abs{ \boldsymbol{k} }$. 
Although reference will be made to scattering in individual partial waves as is natural when close to threshold, we will employ use of scattering quantities constructed directly in real space for the remainder of this work, that most easily connects to the adiabatic approximation asserted in Sec.~\ref{sec:manybody_dynamics}. 
Above, a usual density of states factor is
absorbed into the antisymmetrized (${\cal A}$) and symmetrized (${\cal S}$) momentum eigenstates:
\begin{subequations}
\begin{align}
    \ket{ \boldsymbol{k} }_{\cal A}  
    &=
    i \frac{ \sqrt{ m_r } }{ \pi \hbar }
    \frac{ \sin( \boldsymbol{k} \cdot \boldsymbol{r} ) }{ \sqrt{ 2 } }, \\
    \ket{ \boldsymbol{k} }_{\cal S}  
    &=
    \frac{ \sqrt{ m_r } }{ \pi \hbar }
    \frac{ \cos( \boldsymbol{k} \cdot \boldsymbol{r} ) }{ \sqrt{ 2 } },
\end{align}
\end{subequations}
appropriate to the triplet and singlet scattering channels respectively. That is to say, the states defined above satisfy the energy-normalization condition
\begin{align}
    _{\cal A}\bra{ \boldsymbol{k} }\ket{ \boldsymbol{k}' }_{\cal A} 
    =
    _{\cal S}\bra{ \boldsymbol{k} }\ket{ \boldsymbol{k}' }_{\cal S} 
    =
    \delta(E_k - E'_k) \delta( \phi_k - \phi'_k ),
\end{align}
where $\phi_k$ specifies the incident relative momentum direction.  

As seen in Fig.~\ref{fig:KRb_potentials}, the singlet channel is unfortunately
barrierless in $s$-waves even at the largest electric fields considered by JILA-KRb, permitting molecules to enter the short-range. But even if completely attractive, quantum mechanics still allows nonzero reflection off the dipolar potential in 2D, which certainly already occurs for higher partial waves in the singlet channel. As such, we will also treat the elastic portion of singlet channel scattering in Born approximation. 
This reflection is in contrast to dipolar scattering in three-dimensions, where the $s$-wave scattering amplitude vanishes identically in the Born approximation \cite{Bohn14_PRA}. 

Evaluating the necessary integrals, we obtain the matrix elements analytically as
\begin{widetext}
\begin{subequations} \label{eq:Vint_Born_approximation}
\begin{align}
    _{\cal A}\bra{ \nu; \boldsymbol{k}' }
    \int d z \abs{ \varphi_0(z) }^2
    {V}_{\rm dd}(\boldsymbol{r})
    \ket{ \nu; \boldsymbol{k} }_{\cal A} 
    &= 
    \frac{ k a_{d,\nu} }{ \pi }
    e^{ 
    k^2 a_{\rm ho}^2 
    \xi_{-}(\Delta\phi) 
    } 
    \left[
    \sqrt{ 
    \xi_{+}(\Delta\phi)
    } 
    e^{ k^2 a_{\rm ho}^2 \cos\Delta\phi } 
    {\rm Erfc}\left( 
    k a_{\rm ho} 
    \sqrt{ \xi_{+}(\Delta\phi) }
    \right) 
    \right. \nonumber\\
    &\quad\quad\quad\quad
    \quad\quad\quad\quad
    \quad\left. 
    -
    \sqrt{ 
    \xi_{-}(\Delta\phi)
    } 
    {\rm Erfc}\left(
    k a_{\rm ho} 
    \sqrt{ \xi_{-}(\Delta\phi) }
    \right)
    \right], 
    \\
    _{\cal S}\bra{ \nu; \boldsymbol{k}' }
    \int d z \abs{ \varphi_0(z) }^2
    {V}_{\rm dd}(\boldsymbol{r})
    \ket{ \nu; \boldsymbol{k} }_{\cal S}
    &=  
    \frac{ k a_{d,\nu} }{ \pi }
    \Bigg[
    \frac{ 4 }{ 3 \sqrt{ \pi } }
    \frac{ 1 }{ k a_{\rm ho} }
    -
    e^{ k^2 a_{\rm ho}^2 \xi_{-}(\Delta\phi) }
    \Bigg(
    \sqrt{\xi_{+}(\Delta\phi)} 
    e^{ k^2 a_{\rm ho}^2 \cos\Delta\phi } 
    \text{Erfc}\left( k a_{\rm ho} \sqrt{\xi_{+}(\Delta\phi)} \right)  
    \nonumber\\
    &\quad\quad\quad\quad
    \quad\quad\quad\quad
    \quad\quad\quad\quad
    \quad\quad\quad
    +
    \sqrt{\xi_{-}(\Delta\phi)} 
    \text{Erfc}\left(a_{\rm ho} k \sqrt{\xi_{-}(\Delta\phi)}\right)
    \Bigg)
    \Bigg],
\end{align}
\end{subequations} 
\end{widetext}
where $\xi_{\pm}(\Delta\phi) = ( 1 \pm \cos\Delta\phi ) / 2$ is a function of the scattering angle $\Delta\phi = \cos^{-1} \hat{\boldsymbol{k}} \cdot \hat{\boldsymbol{k}}'$ and ${\rm Erfc}(z)$ is the complementary error function. 
The phase shifts, therefore, also have analytic forms that follow from Eq.~(\ref{eq:Smat_Born_approximation}). 
From the expressions above, we see that the low energy 2D elastic cross section 
scales as $\sigma_{\nu} \propto k (a_{d,\nu})^2$, consistent with Ref.~\cite{Ticknor09_PRA}.
The elastic and inelastic scattering rates are obtained from the relations  
\begin{subequations} \label{eq:collision_rates}
\begin{align}
    \gamma^{\rm el}_{\nu}(\boldsymbol{k})
    &\approx
    n_{\rm 2D}
    \frac{ 2 \pi \hbar }{ m_r }
    \big| 1 - e^{ 2 i \delta_{\nu}(\boldsymbol{k}) } \big|^2, \\
    \gamma^{\rm inel}_{\nu}(\boldsymbol{k})
    &\approx
    n_{\rm 2D} 
    \frac{ 2 \pi \hbar }{ m_r }
    \left(
    1 - \big| e^{ 2 i \delta_{\nu}(\boldsymbol{k}) } \big|^2
    \right), 
\end{align}
\end{subequations}
giving the total scattering rate $\gamma_{\nu}(k) = \gamma^{\rm el}_{\nu}(k) + \gamma^{\rm inel}_{\nu}(k)$, where $n_{\rm 2D}$ is the 2D planar number density. 
The expressions above treat indistinguishable fermions. 

\begin{figure}[ht]
    \centering
    \includegraphics[width=\columnwidth]{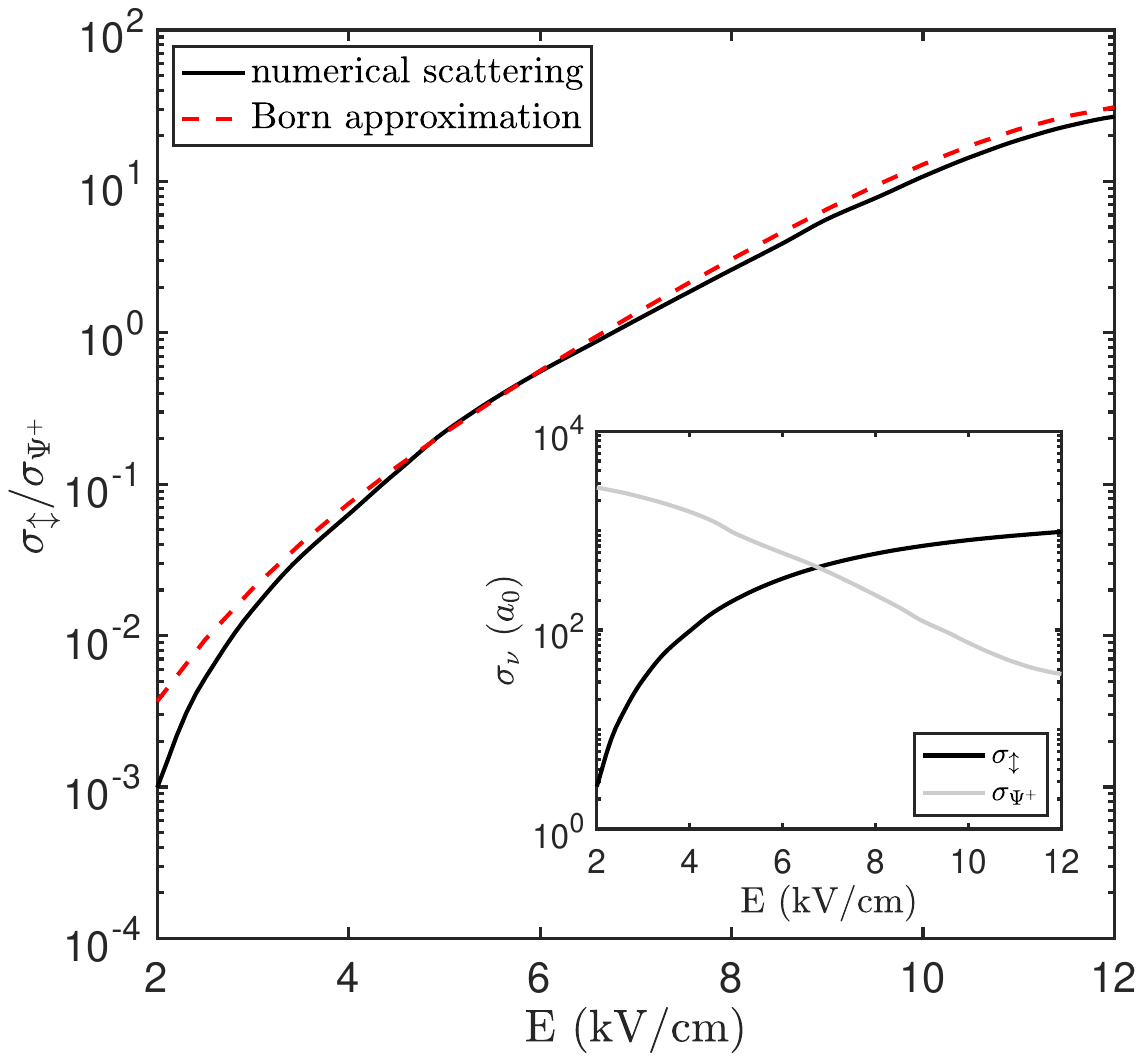}
    \caption{ Ratio of the direct and triplet exchange integral cross sections as a function of electric field, obtained from numerical scattering calculations (red points), and the analytic Born approximation (black curve). The inset plots the numerically computed integral cross sections for direct (black) and triplet exchange (red) interactions as a function of electric field. Collisions occur at $E = 300$ nK. }
    \label{fig:ICSratio_vs_Efield}
\end{figure}

The use of the Born approximation is supported by comparisons of its cross sections with those obtained from numerical scattering calculations. 
We perform the latter via log-derivative propagation of Eq.~(\ref{eq:radial_SE}) with an adaptive-step version of Johnson's algorithm \cite{Johnson73_JCP}, computing the elastic $S$-matrix elements $S_{\nu, \nu}^{m_{\phi}}$. The elastic integral cross sections are then obtained through:
\begin{align} \label{eq:ICS_relation}
    \sigma_{\nu} 
    &=
    \frac{ 2 }{ k_{\nu} } 
    \sum_{m_{\phi}} \abs{ 1 - S_{\nu, \nu}^{m_{\phi}} }^2 \nonumber\\
    &=
    \frac{ 2 \pi }{ k_{\nu} }
    \int_{0}^{2\pi} d(\Delta\phi) 
    \abs{ 1 - e^{ 2 i \delta_{\nu}(k, \Delta\phi) } }^2,
\end{align}
with the factor of 2 assuming indistinguishable scatterers. 
Numerical convergence of scattering calculations is achieved by using up to $m_{\phi} = 11$, where we point out that the 2D cross section for nonzero partial wave scattering is expected to scale as $\sigma_{m_{\phi}} \propto k/m_{\phi}^{4}$ \cite{Ticknor09_PRA}.  
See App.~\ref{app:multichannel_scattering2D} for our adopted convention of scattering quantities. 
In Fig.~\ref{fig:ICSratio_vs_Efield}, we plot $\sigma_{\updownarrow}/\sigma_{\Psi^{+}}$ as a function of electric field as obtained from scattering calculations (red points), and the Born approximation result (black curve).
The data, obtained at a collision energy of $E = 300$ nK, shows excellent agreement with each other.

In ultracold gases of more strongly dipolar molecules, collision times could become comparable to, or even exceed the KDD pulse time intervals. Collisions can then no longer be treated as instantaneous, but must incorporate the periodic time-dependence of the Hamiltonian explicitly.  
This complication is best handled with Floquet theory \cite{Bilitewski15_PRA}, but we leave developments of such spin dynamics to a future work and assume the instantaneous collision regime for the remainder of this paper.

\section{ Many-body itinerant spin dynamics \label{sec:manybody_dynamics} }

We now shift our attention from the dynamics of just two, to many molecules.  
This progression comes naturally since even many-body physics is predominantly emergent from pairwise interactions. 
The Hamiltonian of the $N_{\rm mol}$ molecule system is thus given by
\begin{subequations}
\begin{align}
    H
    &= 
    \sum_{i = 1}^{N_{\rm mol}}
    h_i(t)
    + 
    \sum_{i = 1}^{N_{\rm mol}}
    H_i 
    +
    \sum_{ \langle i, j \rangle }
    V( \boldsymbol{r}_i - \boldsymbol{r}_j ), \\
    H_i
    &=
    \frac{ \boldsymbol{p}_i^2 }{ 2 m }
    + 
    V_{\rm ext}(\boldsymbol{q}_i)
\end{align}
\end{subequations}
where $h_i$ is the Hamiltonian governing the internal states of molecule $i$ (\ref{eq:single_molecule_Hamiltonian}), and $V$ is the molecule-molecule interaction potential with $\langle i, j \rangle$ denoting unique molecular pair indices.

In nondegenerate gases, molecular motion is well described by Maxwell-Boltzmann statistics and classical trajectories. 
We expect the former to be valid when the phase space density of the gas $\rho_{\rm PSD} = n_{\rm 2D} \lambda_{\rm dB}^2$, is much less than unity, where $\lambda_{\rm dB} = \hbar\sqrt{2\pi / (m k_B {\rm T})}$ is the thermal de Broglie wavelength. 
The extent of each molecules wavefunction is then much smaller than the mean spacing between them, so much so that quantum statistical effects are negligible.  
Underlying the latter simplification is the truncated Wigner approximation (TWA) \cite{Polkovnikov10_AP}, that formally starts with the Wigner-Weyl transform that maps operators $\Omega$ living in Hilbert space, into Weyl symbols $\Omega_W$ living in phase space \cite{Wigner32_PR}: 
\begin{align}
    \Omega_W(\boldsymbol{q}, \boldsymbol{p})
    &=
    {\cal W}_{(\boldsymbol{q}, \boldsymbol{p})}[\Omega ] \nonumber\\
    &=
    \int d^3\boldsymbol{q}'
    \bra{ \boldsymbol{q} - \boldsymbol{q}'/2 }
    \Omega 
    \ket{ \boldsymbol{q} + \boldsymbol{q}'/2 }
    e^{ \frac{ i }{ \hbar } \boldsymbol{p} \cdot \boldsymbol{q} }. 
\end{align} 
The Weyl symbol corresponding to the reduced density matrix for the motion of a single molecule $W(\boldsymbol{q}, \boldsymbol{p}) = {\cal W}_{(\boldsymbol{q}, \boldsymbol{p})}[\rho]$, known as the Wigner function, represents the molecular distribution in quantum phase space. Time evolution of $W(\boldsymbol{q}, \boldsymbol{p})$ \cite{Moyal49_MPCPS, Baker58_PR} is, under the TWA, given by the Boltzmann equation: 
\begin{align} \label{eq:Boltzmann_equation}
    \left(
    \frac{ \partial }{ \partial t }
    +
    \frac{ \boldsymbol{p} }{ m } \cdot 
    \frac{ \partial }{ \partial \boldsymbol{q} }
    -
    \frac{ \partial V_{\rm ext} }{ \partial \boldsymbol{q} } \cdot  
    \frac{ \partial }{ \partial \boldsymbol{p} }
    \right)
    W(\boldsymbol{q}, \boldsymbol{p})
    =
    {\cal I}[ W ],
\end{align}
where ${\cal I}[ W ]$ is the collision integral that accounts for all two-body scattering events. 

As a result, the TWA approximates the expectation value over external DoF of an operator $\Omega$, that may have support on both internal and external DoF, in time as 
\begin{align} \label{eq:motion_averaged_Heisenberg_operators}
    \Omega(t)
    &=
    \langle \Omega(\boldsymbol{q}, \boldsymbol{p}; t) \rangle_{W_0} \nonumber\\
    &\approx 
    \int d^3\boldsymbol{q} d^3\boldsymbol{p}
    W[ \boldsymbol{q}(0), \boldsymbol{p}(0) ] 
    \Omega_W[ \boldsymbol{q}_{\rm cl}(t), \boldsymbol{p}_{\rm cl}(t) ],
\end{align}
where 
$\{ \boldsymbol{q}_{\rm cl}(t), \boldsymbol{p}_{\rm cl}(t) \}$ denotes classical trajectories in phase space, 
and $\langle \ldots \rangle_{W_0}$ is an ensemble average over $W[ \boldsymbol{q}(0), \boldsymbol{p}(0) ]$. 
The use of classical trajectories immediately implies that external molecular motion is treated as an incoherent process without coherent coupling to internal molecular dynamics, 
an approximation afforded by barriered collisions as described in Sec.~\ref{sec:scattering_theory}.
Integrating over a classical distribution $W_0$ can, therefore, be treated as averaging multiple disorder realizations of the quantum spin dynamics.

With the internal molecular states involving much larger energy scales ($\sim$GHz) than the molecular motion ($\sim$kHz), utilizing classical trajectories can also be seen as a form of adiabatic approximation with $\{\boldsymbol{q}, \boldsymbol{p}\}$ constituting the slow DoF. 
In the event where two collided molecules proceed to make unhindered excursions around the trap only to re-collide with one another, their combined two-body state will develop a Berry phase due to the closed-loop formed by their adiabatic coordinates $\boldsymbol{r} = \boldsymbol{q}_1 - \boldsymbol{q}_2$. 
We will assume that such occurrences are extremely rare in a dilute gas, so we ignore such geometric phases here completely.

During a collision, the interactions give rise to entanglement between the two-molecules \cite{Jaksch99_PRL, Brennen00_PRA}, transforming the two-molecule internal state and setting the outbound molecules off on a collision-modified trajectory.  
The post-collision trajectories are determined by the $S$-matrix $S_{\nu', \nu}(\hat{\boldsymbol{k}}, \hat{\boldsymbol{k}}')$, with joint probability distribution of the outbound and inbound scattering directions given by
\begin{align} \label{eq:scattering_distribution}
    \mathbb{P}(\hat{\boldsymbol{k}}', \hat{\boldsymbol{k}})
    &=
    \tr_{\nu}\left\{ 
    S(\hat{\boldsymbol{k}}, \hat{\boldsymbol{k}}') 
    \rho
    S^{\dagger}(\hat{\boldsymbol{k}}, \hat{\boldsymbol{k}}')
    \right\},
\end{align}
where $\rho$ is the two-molecule reduced density matrix, and the trace above runs over all two-body scattering channels $\nu$.   
Having the $S$-matrix constructed with continuous coordinates $\hat{\boldsymbol{k}}$ and $\hat{\boldsymbol{k}}'$, makes explicit that spin and motion will remain only classically correlated as follows from the adiabatic approximation.

Although each collision results in a local interaction, the itinerance of the molecules can give rise to highly nonlocal connectivity over time as they interact freely with various partners.   
This effective all-to-all interaction could lead to interesting quantum multi-body processes, the physics of which we now turn our attention to.

\subsection{ Markovian simulations for JILA-KRb \label{sec:Markov_simulations} }

In a dilute gas of a large number of molecules, as is the case in JILA-KRb, it is highly unlikely that two colliding molecules would ever re-encounter the same collision partner. 
This assumption is especially valid in the presence of large $s$-wave losses, given that molecules which get entangled tend to be lost from subsequent collisions (further discussed in App.~\ref{app:collision_stats} and Sec.~\ref{sec:loss_autoselection} below). 
At the level of the one-body reduced density matrix, this assumption implies that the full many-body dynamics will be indistinguishable from a scenario where each molecule only ever encounters new (i.e. not previously encountered) collision partners. 
Importantly, however, these new partners could have themselves undergone prior collisions with other molecules as well. 
We will handle the dynamics of this regime in a Markovian fashion, where the one-body reduced density matrix of a molecule is determined only by their most recent collision.

With the scattering channel-dependent phase shifts computed in Eq.~(\ref{eq:Vint_Born_approximation}) of Sec.~\ref{sec:scattering_phaseshifts}, we now explicitly derive Kraus operators on the single-molecule spin space, that follow from tracing out one molecule from the two-molecule density matrix following a collision.
Represented in the symmetrized basis (\ref{eq:2molecule_basis}), a collision imparts scattering phase shifts to each of the appropriately symmetrized states of the two-body density matrix:
\begin{align}
    \rho'
    &=
    S 
    \rho 
    S^{\dagger} 
    =
    \sum_{\mu, \nu}
    e^{2 i \delta_{\mu}}
    \ket{ \mu }
    \rho_{\mu, \nu}
    \bra{ \nu }
    e^{-2 i \delta_{\nu}}.
\end{align}
Expressed as an operator acting on the product basis states $\{ \ket{\downarrow\downarrow}, \ket{\downarrow\uparrow}, \ket{\uparrow\downarrow}, \ket{\uparrow\uparrow} \}$, the $S$-matrix is then given a matrix representation: 
\begin{align} \label{eq:2body_gate_matrix} 
    \boldsymbol{S}
    &= 
    \begin{pmatrix}
        e^{2 i \delta_{\updownarrow}}
        & 0 & 0 & 0 
        \\
        0 & \frac{ e^{2 i \delta_{\Psi^{+}}} + e^{2 i \delta_{\Psi^{-}}} }{ 2 } & \frac{ e^{2 i \delta_{\Psi^{+}}} - e^{2 i \delta_{\Psi^{-}}} }{ 2 } & 0
        \\
        0 & \frac{ e^{2 i \delta_{\Psi^{+}}} - e^{2 i \delta_{\Psi^{-}}} }{ 2 } & \frac{ e^{2 i \delta_{\Psi^{+}}} + e^{2 i \delta_{\Psi^{-}}} }{ 2 } & 0  
        \\
        0 & 0 & 0 & e^{2 i \delta_{\updownarrow}} 
    \end{pmatrix}.
\end{align}
The structure of the matrix above indicates that it commutes with $\Sigma_Z = \sum_{i=1}^{N_{\rm mol}} \sigma_{Z}^{(i)}$, where $\sigma_{Z}^{(i)}$ is the $Z$ Pauli matrix on molecule $i$, resulting in a U(1) conservation law of $\Sigma_Z$ charge.  
The result is that Bloch vectors in the azimuthal plane will never leave the plane under application of these gates.

The resulting post-collision reduced density matrix of molecule $A$ is then obtained by taking a partial trace over molecule $B$ 
\begin{align}
    \boldsymbol{\varrho}'_A
    &=
    \tr_B\{ 
    \boldsymbol{S} 
    \boldsymbol{\rho}
    \boldsymbol{S}^{\dagger} 
    \}, 
\end{align} 
that will also be the reduced density of molecule $B$ since the molecules are identical. 
Keeping track of only the one-body reduced density matrices allows us to efficiently simulate the Markovian itinerant spin dynamics with a modified direct simulation Monte Carlo method \cite{Bird70_PF}. Although details of the simulation are already outlined in the Supplementary Materials of Ref.~\cite{Carroll25_Sci}, we present an abridged version here once more for completeness of discussions in this paper.

\subsubsection{ Monte Carlo simulations }

In our mixed quantum-classical treatment of Markovian spin dynamics, the simulation commences by taking $W[\boldsymbol{q}(0), \boldsymbol{q}(0)]$ to be a Maxwell-Boltzmann distribution, and approximating it with discrete phase space points sampled from it:
\begin{align} \label{eq:discrete_Wigner}
    W[\boldsymbol{q}(0), \boldsymbol{q}(0)]
    &\approx 
    \sum_{i=1}^{N_{\rm mol}}
    \delta^2( \boldsymbol{q} - \boldsymbol{q}_i )
    \delta^2( \boldsymbol{p} - \boldsymbol{p}_i ).
\end{align} 
All the molecules are assumed identically prepared in the $\ket{ + }$ state. 
The molecules then undergo classical motion in phase space, progressing forward in discrete time steps of $\Delta t$ via St\"ormer-Verlet symplectic integration \cite{Verlet67_PR}.
Alongside the molecular positions $\boldsymbol{q}_i$ and momenta $\boldsymbol{p}_i$, we also keep track of the one-body reduced density matrix $\boldsymbol{\varrho}$ of each molecule, but ignore all bookkeeping of quantum correlations between molecules  
\footnote{
Tracking only the single-molecule states is valid since subsequent collisions of a molecule $B$, previously collisionally entangled with another molecule $A$, cannot decrease the reduced density matrix purity of $A$ so long as the subsequent collision partners of $B$, or $B$ itself, do not re-collide with $A$, which is the Markov approximation. More details can be found in the Supplementary Materials of Ref.~\cite{Beckey21_PRL}.
}.

Collisions are sampled using the (DSMC) method \cite{Bird70_PF, Wang20_PRA}, which exploits the locality of interactions for computational efficiency. 
In our implementation, the simulation volume is first partitioned into discrete grid cells of volume $\Delta V_{\rm cell}$, into which the simulated molecules are binned based on their positions. Collisions are then assumed to only occur within each grid cell with probability
\begin{align} \label{eq:collision_probability}
    P_{\rm coll}(k)
    &=
    \frac{ \Delta t }{ \Delta V_{\rm cell} }
    \sum_{\nu}
    \rho_{\nu, \nu}
    \gamma_{\nu}(k),
\end{align}
which depends on the relative momentum $\hbar |\boldsymbol{k}| = |\boldsymbol{p}_A - \boldsymbol{p}_B|$, and the 2-body reduced density matrix $\boldsymbol{\rho} = \boldsymbol{{\cal U}} \boldsymbol{\varrho}_A \otimes \boldsymbol{\varrho}_B \boldsymbol{{\cal U}}^{\dagger}$ transformed into the appropriately symmetrized basis (\ref{eq:2molecule_basis}) by $\boldsymbol{{\cal U}}$.  
If determined to occur, the collision must be assigned as elastic or inelastic, done as follows. 

Given that all $p$-wave ($\abs{m_{\phi}} = 1$) losses are far suppressed over $s$-wave ($m_{\phi} = 0$) ones (see Sec.~\ref{sec:scattering_phaseshifts}), we will treat all scattering phase shifts in the symmetric sector as real-valued.  
As for the antisymmetric singlet channel, scattering of identical fermions in $\ket{\Psi^{-}}$ must necessarily involve the $m_{\phi} = 0$ partial wave, which results in short-range inelastic loss.
We model scattering in the antisymmetric sector with a complex phase shift $\delta_{\rm anti} = \delta_{\Psi^{-}} + i \eta_s$, comprising a real-valued dipolar part $\delta_{\Psi^{-}}$ and a purely imaginary $s$-wave contribution $i\eta_s$. 
The imaginary $s$-wave phase shift is motivated by the experimentally observed universal short-range loss in KRb \cite{Ni10_Nat, Quemener10_PRA}.
By careful comparison with the experimentally measured number loss rates, we insert an empirically determined imaginary phase shift of $\eta_s = 0.05$ that implies about a $20 \%$ probability of loss for collisions in the singlet channel.  
The resultant inelastic scattering rate is then computed as
\begin{align}
    \gamma_{\Psi^{-}}^{\rm inel}
    &\approx 
    n_{\rm 2D}
    \frac{ 4 \pi \hbar }{ \mu }
    \left(
    1
    -
    e^{ -4 \eta_s }
    \right), 
\end{align}
so that the total scattering rate in channel $\ket{\Psi^{-}}$ is given as $\gamma_{\Psi^{-}} = \gamma^{\rm el}_{\Psi^{-}} + \gamma^{\rm inel}_{\Psi^{-}}$. The probability that a simulated collision occurrence is inelastic is then:
\begin{align}
    \mathbb{P}_{\rm inel}
    &=
    \frac{ \rho_{\Psi^{-}, \Psi^{-}} \gamma_{\Psi^{-}}^{\rm inel} }{
    \sum_{\nu}
    \rho_{\nu, \nu}
    \gamma_{\nu}
    }.
\end{align}
Inelastic collisions are exothermic and result in trap-loss of the molecular pair, translating to a discarding of these molecules from the Monte Carlo simulation.

\begin{figure}[ht]
    \centering
    \includegraphics[width=\columnwidth]{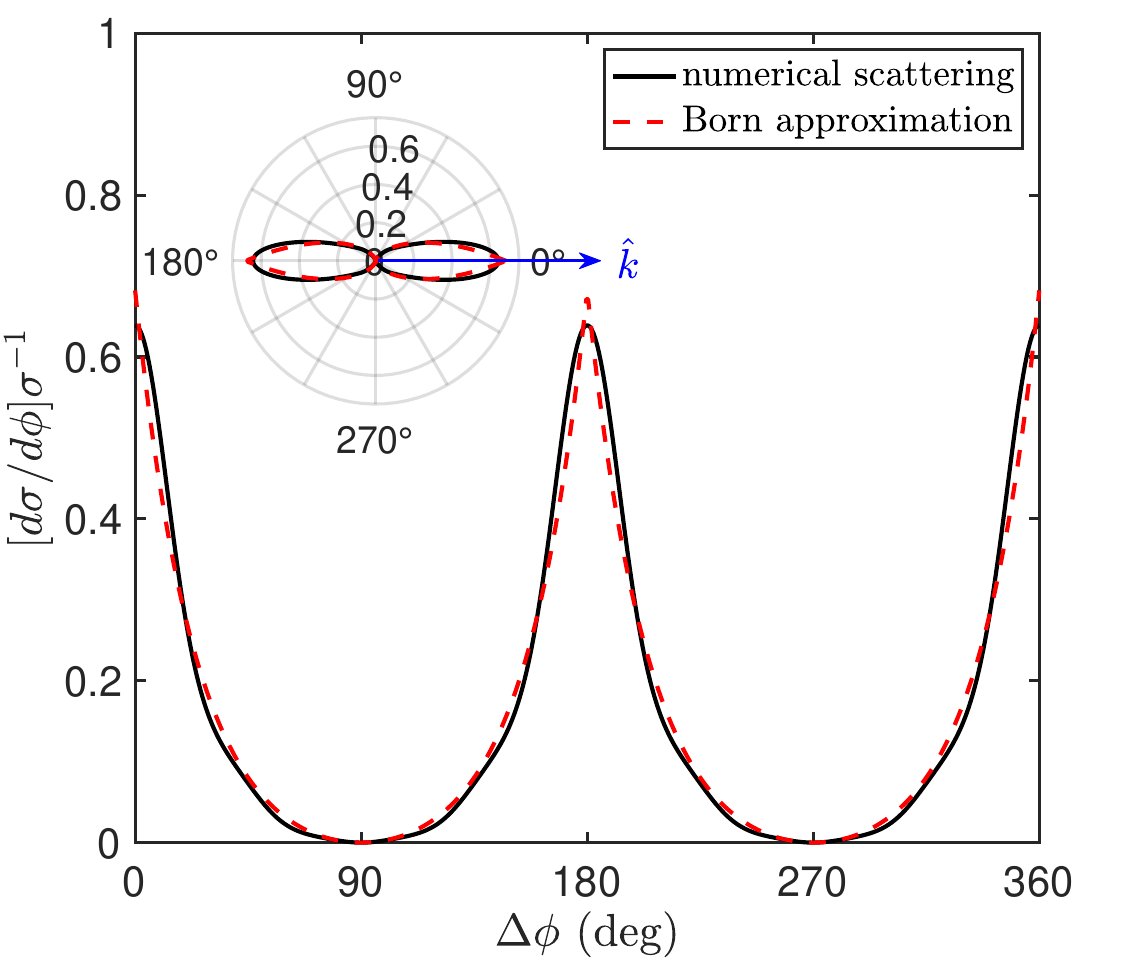}
    \caption{ Direct interacting differential cross section, normalized by the total cross section, as a function of the scattering angle $\Delta\phi$, from numerical scattering calculations (solid black curve) and the Born approximation (dashed red curve) at an applied electric field of ${\rm E} = 12$ kV/cm and collision energy of $E/k_B = 300$ nK. The top-left inset is a polar plot of the same quantity, with the blue arrow indicating the incident scattering direction $\hat{\boldsymbol{k}}$. }
    \label{fig:KRb_DCS_Ef12_fz10k}
\end{figure}

If elastic, the collision must modify the reduced density matrix. 
First, the nondegenerate temperatures of the experiment result in a strong preference for forward/backward scattering, showcased in an exemplary plot of the differential scattering cross section as a function of scattering angle $\Delta\phi$ in Fig.~\ref{fig:KRb_DCS_Ef12_fz10k}. 
The figure gives the cross section obtained from both the Born approximation (dashed red curve) and numerically (solid black curve). This angular character allows us to approximate differential scattering by sampling the scattering angle as $\Delta\phi = 0$ or $\pi$  with equal probability, to produce $\boldsymbol{k}' = +\boldsymbol{k}$ or $-\boldsymbol{k}$ respectively.
Then taking the symmetrized 2-body density matrix, we apply only the elastic scattering phase shifts to it $\Re\{ \delta_{\nu} \}$.

In JILA-KRb, measurements of the Ramsey contrast $C(t)$ are performed, that are equivalent to single-molecule measuring the length of the Bloch vector $\boldsymbol{\sigma}$, with components $\sigma_i$ being Pauli matrices along each Bloch sphere axis $i \in \{ X, Y, Z \}$. 
More details of the experiment can be found in Ref.~\cite{Carroll25_Sci}. 
Leveraging the symmetries in Eq.~(\ref{eq:2body_gate_matrix}), we can extract the contrast within our simulation as the expectation of $\sigma_X$, with respect to each single-molecule density matrix $\langle \sigma_X \rangle_{\varrho(t)} = \tr\{ \sigma_X \varrho(t) \}$. The ensemble averaged contrast is then obtained by taking the mean value of $\langle \sigma_X \rangle_{\varrho(t)}$ over all simulated molecules at any given time $t$, further averaged over multiple simulation shots:
\begin{align}
    C(t)
    &\approx 
    \sum_{i=1}^{N_{\rm mol}}
    \tr\{ 
    \sigma_X \boldsymbol{\varrho}_i[ \boldsymbol{q}_i(t), \boldsymbol{p}_i(t) ] 
    \}.
\end{align}
This expression is exactly what one arrives at after plugging Eq.~(\ref{eq:discrete_Wigner}) into Eq.~(\ref{eq:motion_averaged_Heisenberg_operators}), and taking the expectation of $\Omega = \sigma_X$ where $\{ \boldsymbol{q}_i(t), \boldsymbol{p}_i(t) \}$ are classical trajectories with initial conditions sampled from $W[\boldsymbol{q}(0), \boldsymbol{p}(0)]$.

\begin{figure}[ht]
    \centering
    \includegraphics[width=\columnwidth]{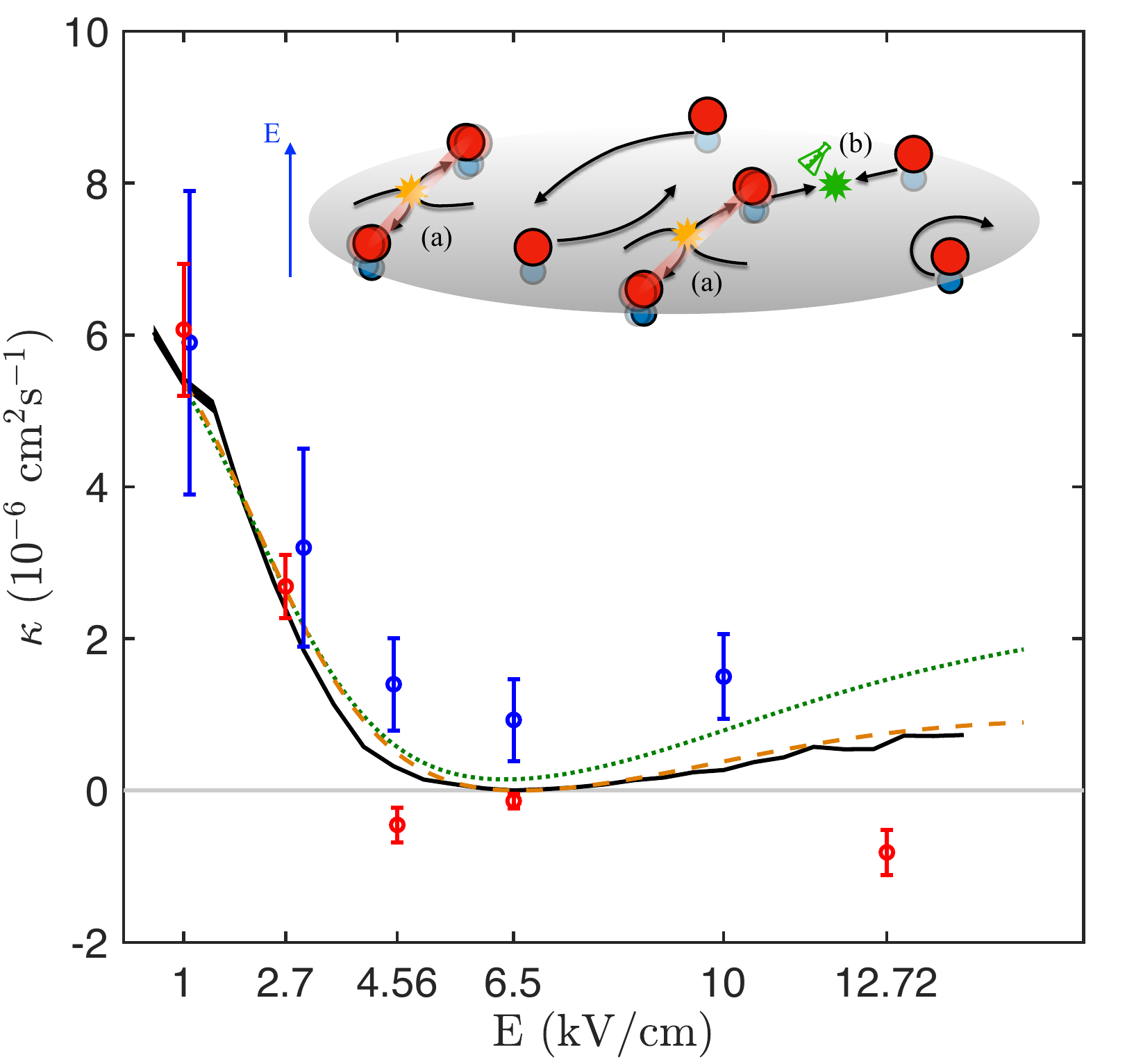}
    \caption{ Density normalized contrast decay rate $\kappa$ as a function of electric field ${\rm E}$, as measured by JILA-KRb (red points with error bars from Ref.~\cite{Carroll25_Sci}) and compared to our analytic theory (orange dashed line) and Monte Carlo simulations (black shaded curve). 
    Comparison is also made to the Floquet engineered JILA experiment (blue points with error bars from Ref.~\cite{Miller24_Nat}), and the case where dynamical decoupling is not effective to average direct interactions (dotted green line). 
    The inset shows an illustration of itinerant molecules and the various forms of collisional interactions they encounter: (a) elastic entangling collisions, and (b) inelastic or chemically reactive collisions that result in trap loss. }
    \label{fig:contrastDecayRate_vs_electricField}
\end{figure}

From time traces of the Ramsey contrast, we extract a density normalized contrast decay rate $\kappa$ by fitting each time trace to a stretched exponential:
\begin{align} \label{eq:stretched_exponential}
    C_{\rm fit}(t)
    &=
    e^{ -( \Gamma t )^{\zeta} },
\end{align}
where $\Gamma$ is the dephasing rate and $\zeta$ is the stretching parameter, 
then fitting a linear density trend to the various dephasing rates
\begin{align} \label{eq:linear_contrastDecayDensityTrend}
    \Gamma(n_{\rm 2D}) 
    &=
    \Gamma_0 + \kappa n_{\rm 2D},    
\end{align}
with background dephasing rate $\Gamma_0$.
A stretched exponential, suspected to be relevant in systems with glassy dynamics \cite{Signoles21_PRX}, is used here for arguments that will be made in Sec.~\ref{sec:loss_autoselection}.   
The extraction procedure above is repeated for various values of the electric field to yield the plot of $\kappa$ vs ${\rm E}$ in Fig.~\ref{fig:contrastDecayRate_vs_electricField}, comparing $\kappa$ taken from JILA-KRb (red points with error bars) and Monte Carlo simulations (black curve with the surrounding shaded region denoting Monte Carlo sampling error bars). 
$\zeta$ is found to be less than unity for all values of ${\rm E}$.
Both quantitative and qualitative trend agreement between the theory and experiment of $\kappa$ versus ${\rm E}$, indicate that our theory incorporates the essential physics behind the Ramsey contrast dynamics. 
We also provide a plausible explanation for the points that disagree later in Sec.~\ref{sec:loss_autoselection}.

At an electric field of ${\rm E} \approx 6.5$ kV/cm, the Ramsey contrast decay is seen to come to a complete halt in both the theory and experiment. 
To understand what underlies this protection of spin coherence, 
it is convenient to decompose the scattering $S$-matrix (\ref{eq:2body_gate_matrix}) into two-site Pauli operators $\{ \sigma_X, \sigma_Y, \sigma_Z \}$, which we determine to be 
\begin{align} \label{eq:2body_gate_representation}
    \boldsymbol{S}
    &=
    \frac{ g_0 }{ 4 }
    \mathbb{I} \otimes \mathbb{I} 
    +
    \frac{ g_1 }{ 4 }
    \sigma_Z \otimes \sigma_Z 
    \nonumber\\
    &\quad
    +
    \frac{ g_2 }{ 4 }
    \left[
    \sigma_X \otimes \sigma_X + \sigma_Y \otimes \sigma_Y + \sigma_Z \otimes \sigma_Z
    \right], 
\end{align} 
where 
\begin{subequations}
\begin{align}
    g_0 
    &=
    2 e^{2 i \delta_{\updownarrow}} + e^{2 i \delta_{\Psi^{-}}} + e^{2 i \delta_{\Psi^{+}}}, \\
    g_1
    &=
    2 ( e^{2 i \delta_{\updownarrow}} - e^{2 i \delta_{\Psi^{+}}} ), \\
    g_2
    &=
    e^{2 i \delta_{\Psi^{+}}} - e^{2 i \delta_{\Psi^{-}}},
\end{align}
\end{subequations}
are the two-site Pauli operator coefficients. 
It is now evident that when ${\rm E}$ is tuned so that $\delta_{\updownarrow} = \delta_{\Psi^{+}}$, the operator reduces to $\boldsymbol{S} = ({ g_0 / 4 }) \mathbb{I} + ({ g_2 / 4 }) \left[ \sigma_X \sigma_X + \sigma_Y \sigma_Y + \sigma_Z \sigma_Z \right]$,
dropping the explicit tensor products for notational brevity.
One can show that the eigenstates of this operator are in fact all two-spin singlet and triplet states (\ref{eq:2molecule_basis}), with the states in each sector being degenerate. The $\ket{+} \otimes \ket{+}$ state is, therefore, an eigenstate of this operator and remains stationary under its application.  
This special value of the electric field is referred to as the \textit{Heisenberg point} with dynamics generated by $\boldsymbol{\sigma} \cdot \boldsymbol{\sigma}$ type interactions, rendering the entire two-molecule triplet sector and all possible superpositions of tensor products with these pair states, a decoherence free subspace \cite{Lidar14_JWS}.

\subsubsection{ Quantum Markov chain model }

To gain further intuition of the contrast dynamics, we consider a scenario in which the ultracold gas remains dilute while taking $N_{\rm mol} \rightarrow \infty$. In these limits, we can approximate the dephasing of one-body density matrices as caused by collisional encounters with partners strictly in the $\varrho(0) = \ket{+}\bra{+}$ state.
Such a series of encounters constitutes a quantum Markov chain process.
In the parlance of quantum Shannon theory, we identify the so-called collision map \cite{Ciccarello22_PR}:
\begin{align}
    {\cal E}[\boldsymbol{\varrho}_A] 
    &=
    \sum_{ \alpha_B = \downarrow, \uparrow }
    \left( \boldsymbol{\Pi}_{\alpha_B}  
    \boldsymbol{S} \right)
    \left[ \boldsymbol{\varrho}_A \otimes \boldsymbol{\varrho}_B(0) \right]
    \left( \boldsymbol{\Pi}_{\alpha_B}
    \boldsymbol{S} \right)^{\dagger} \nonumber\\
    &=
    \sum_{ \alpha = \downarrow, \uparrow }
    \boldsymbol{K}_{\alpha}
    \boldsymbol{\varrho}_A
    \boldsymbol{K}_{\alpha}^{\dagger}, 
\end{align}
as a noisy quantum channel \footnote{ Note that the use of channel here differs from that employed to describe scattering channels, but is rather a channel in the quantum information theory sense [cite Nielson and Chuang]. }, with projectors
\begin{subequations}
\begin{align}
    \boldsymbol{\Pi}_{\downarrow_B}
    &=
    \begin{blockarray}{ccccc}
        & 
        \bra{\downarrow_B}\ket{\downarrow\downarrow} & 
        \bra{\downarrow_B}\ket{\downarrow\uparrow} &
        \bra{\downarrow_B}\ket{\uparrow\downarrow} & 
        \bra{\downarrow_B}\ket{\uparrow\uparrow}   \\
        \begin{block}{c(cccc)}
        \ket{\downarrow_A}\:\: & 1 & 0 & 0 & 0 \\
        \ket{\uparrow_A}\:\: & 0 & 0 & 1 & 0 \\
        \end{block}
    \end{blockarray}\:\:, \\
    \boldsymbol{\Pi}_{\uparrow_B}
    &= 
    \begin{blockarray}{ccccc}
        & 
        \bra{\uparrow_B}\ket{\downarrow\downarrow} & 
        \bra{\uparrow_B}\ket{\downarrow\uparrow} &
        \bra{\uparrow_B}\ket{\uparrow\downarrow} & 
        \bra{\uparrow_B}\ket{\uparrow\uparrow}   \\
        \begin{block}{c(cccc)}
        \ket{\downarrow_A}\:\: & 0 & 1 & 0 & 0 \\
        \ket{\uparrow_A}\:\: & 0 & 0 & 0 & 1 \\
        \end{block}
    \end{blockarray}\:\:,
\end{align}
\end{subequations}
and the Kraus operators $\boldsymbol{K}_{\alpha} = \boldsymbol{\Pi}_{\alpha_B} \boldsymbol{S} \left( \mathbb{I}_A \otimes \ket{ + } \right)$ with $\alpha=\downarrow,\uparrow$. 
We can write the Kraus operators explicitly as:
\begin{subequations}
\begin{align}
    \boldsymbol{K}_{\downarrow}
    &=
    \frac{1}{\sqrt{2}}
    \begin{pmatrix}
        e^{2 i \delta_{\updownarrow}}
        & 0 \\
        \frac{ e^{2 i \delta_{\Psi^{+}}} - e^{2 i \delta_{\Psi^{-}}} }{ 2 } & \frac{ e^{2 i \delta_{\Psi^{+}}} + e^{2 i \delta_{\Psi^{-}}} }{ 2 }
    \end{pmatrix}, \\
    \boldsymbol{K}_{\uparrow}
    &= 
    \frac{1}{\sqrt{2}}
    \begin{pmatrix}
        \frac{ e^{2 i \delta_{\Psi^{+}}} + e^{2 i \delta_{\Psi^{-}}} }{ 2 } & \frac{ e^{2 i \delta_{\Psi^{+}}} - e^{2 i \delta_{\Psi^{-}}} }{ 2 } \\
        0 & e^{2 i \delta_{\updownarrow}} 
    \end{pmatrix},
\end{align}
\end{subequations}
that are easily shown to satisfy $\sum_{ \alpha } \boldsymbol{K}_{\alpha}^{\dagger} \boldsymbol{K}_{\alpha} = \mathbb{I}$. 

The subsequent derivation considers only the one-body density matrix, so we will drop any label of molecule $A$ or $B$ in what follows. 
Treated in terms of discrete collision instances, the update rule for the one-body density matrix is given by $\boldsymbol{\varrho}' = {\cal E}[\boldsymbol{\varrho}] = \sum_{ \alpha } \boldsymbol{K}_{\alpha} \varrho \boldsymbol{K}_{\alpha}^{\dagger}$.
Then for an initial molecule in state $\ket{+}\bra{+}$, the post-collision reduced density matrix of the molecule after a single collision is given by
\begin{align}
    \boldsymbol{\varrho}'
    &=
    \frac{1}{2}
    \begin{pmatrix}
        1 & 
        \cos[ 2 (\delta_{\Psi^{+}} - \delta_{\updownarrow}) ]
        \\
        \cos[ 2 (\delta_{\Psi^{+}} - \delta_{\updownarrow}) ]
        & 1
    \end{pmatrix},
\end{align}
which gives the post-collision Ramsey contrast in terms of phase shifts as
$\langle \sigma_X \rangle_{\varrho'} = \tr\{ \sigma_X \boldsymbol{\varrho}' \} = \cos[ 2 (\delta_{\Psi^{+}} - \delta_{\updownarrow}) ]$. 
The change in contrast after a single collision therefore evaluates to 
\begin{align} \label{eq:change_in_contrast}
    \Delta C
    &=
    1 - \langle \sigma_X \rangle_{\varrho'} 
    =
    2 \sin^2\left(
    \delta_{\Psi^{+}} - \delta_{\updownarrow}
    \right).
\end{align}

We thus infer an early-time contrast decay rate by taking it as the linear slope over which the contrast changes within the time interval $\Delta t = 1/\gamma_{\rm el}$:
\begin{align} \label{eq:contrast_decay_rate}
    \Gamma 
    &\approx
    \frac{ \Delta C }{ \Delta t } 
    =
    \gamma_{\rm el} \Delta C, 
\end{align}
where $\gamma_{\rm el} = \left( \gamma_{\updownarrow} + \gamma_{\Psi^{+}} \right) / 2$.
Since still dependent on $k$, it is more appropriate to consider a thermally averaged contrast decay rate $\langle \Gamma \rangle_{W_0}$, obtained by integrating $\gamma_{\rm el}(k)$ and $\Delta C(k)$ over the equilibrium Maxwell-Boltzmann distribution individually, then taking their product. 
We plot $\kappa \approx \gamma_{\rm el} \Delta C / n_{\rm 2D}$ as a dashed orange curve in Fig.~\ref{fig:contrastDecayRate_vs_electricField}, which already gives excellent agreement with the experimental observations, and even more so with full Monte Carlo simulations. 

We also determine the validity of averaged direct interactions in Eq.~(\ref{eq:direct_dipole_length}) from dynamical decoupling pulses. 
In the case where the phase shifts accrued in the scattering channels $\ket{ \Uparrow }$ and $\ket{ \Downarrow }$ take their native values, we find that the change in contrast after a single collision is instead
\begin{align} \label{eq:contrastchange_noDD}
    \Delta C
    &=
    1
    -
    \cos\left[
    2 \delta_{\Psi^{+}} - ( \delta_{\Downarrow} + \delta_{\Uparrow} )
    \right]
    \cos( \delta_{\Downarrow} - \delta_{\Uparrow} ).
\end{align}
The relevant collision rate is then taken to be $\gamma_{\rm el} = \left( \gamma_{\Downarrow} + \gamma_{\Uparrow} + 2 \gamma_{\Psi^{+}} \right) / 4$, for which the density normalized thermally averaged contrast decay rate is plotted as the dotted green curve in Fig.~\ref{fig:contrastDecayRate_vs_electricField}. 
Although little difference is observed at low field, we find that inclusion of the differential phase shift between $\ket{ \Uparrow }$ and $\ket{ \Downarrow }$ states results in the absence of a Heisenberg point, supporting use of Eq.~(\ref{eq:direct_dipole_length}).  
In addition, we also plot data (blue points with error bars) from a related experiment where the intermolecular interactions were attained through Floquet engineering instead of a static field \cite{Miller24_Nat}. 
The Floquet pulses used were spaced by $100$ $\mu$s, longer than the KDD pulses in Ref.~\cite{Carroll25_Sci}, leading to a contrast decay trend suggestive of that predicted by Eq.~(\ref{eq:contrastchange_noDD}) where dynamical decoupling fails.  
However, the exact mechanism for increased contrast decay in these latter experiments remains unclear, with other possible explanations such as inelastic Floquet scattering \cite{Bilitewski15_PRA, Karman25_arxiv} warranting further future investigations.

\subsection{ Loss-induced autoselection of pure states \label{sec:loss_autoselection} }

Although able to capture the short-time decay of Ramsey contrast, we find that beyond first-encounter two-body collisions are ultimately necessary to capture the full Ramsey contrast dynamics at long times.   
In particular, collisional loss in the molecular gas can significantly suppress the collective long-time Ramsey contrast decay.
This suppression is a spin-motion coupled effect, where subsequent collisions after the first can experience scattering in the singlet-channel, involving $s$-wave collisional angular momentum from Fermi symmetry. 
The introduction of the singlet channel can be understood from the following process: a first collision occurs between two molecules $A$ and $B$ with initial states $\varrho_A(0) = \varrho_B(0) = \ket{ + }\bra{ + }$. Upon its second collisional encounter with another molecule $C$ in state $\varrho_C(0) = \ket{ + }\bra{ + }$, the joint pre-collision state of molecule $C$ and the now decohered $A$ is given by $\rho = {\cal E}[ \varrho_A(0) ] \otimes \varrho_C(0)$. This two-body product state is no longer confined within the triplet subspace (\ref{eq:2molecule_basis}), but now has a nonzero singlet state component given by:
\begin{align}
    \langle \Psi^{-} | \rho | \Psi^{-} \rangle
    &=
    \frac{1}{2} \sin^2( \delta_{\updownarrow} - \delta_{\Psi^{+}} ).
\end{align}   

\begin{figure}[ht]
    \centering
    \includegraphics[width=\columnwidth]{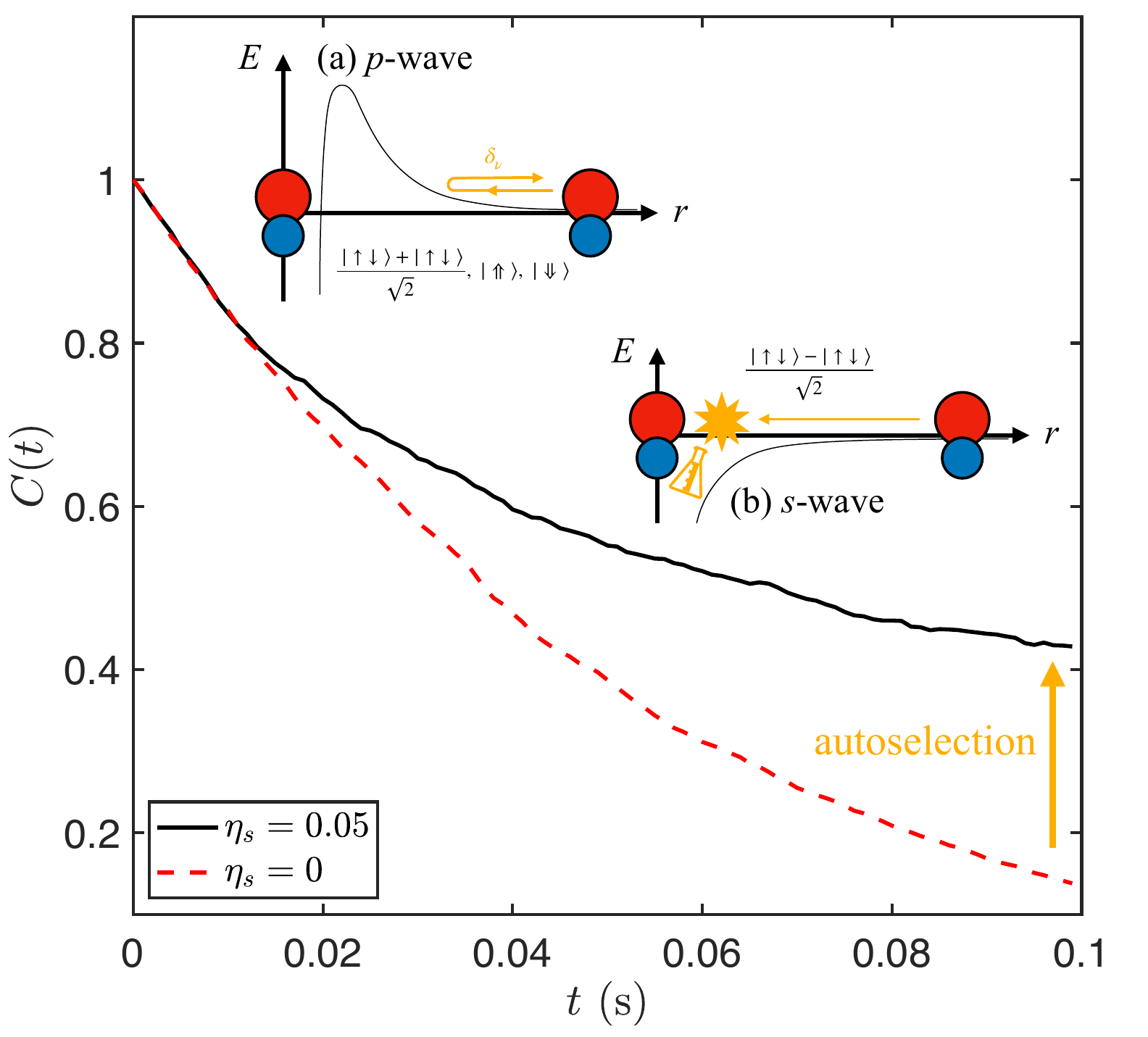}
    \caption{ Contrast decay as a function of time from Monte Carlo simulations including loss (solid black curve with $\eta_s = 0.05$) and no loss (dashed red curve with $\eta_s = 0$). The simulation treats a gas of $N_{\rm mol} = 400$ molecules at temperature ${\rm T} = 300$ nK, with applied electric field $E = 12.7$ kV/cm. 
    The insets illustrate (a) barriered collisions in the triplet sector and (b) barrierless collisions in the singlet sector.
    }
    \label{fig:contrast_loss_vs_noloss}
\end{figure}

As seen in the inset of Fig.~\ref{fig:KRb_potentials}\textcolor{blue}{a}, the molecules have attractive dipole-dipole interactions in the singlet channel and, moreover, have no collisional $p$-wave potential barrier to prevent entry into the short range as illustrated by inset (b) in Fig.~\ref{fig:contrast_loss_vs_noloss}.
In turn, decohered molecules have a higher tendency to undergo chemical reactions and be lost from the trap, which erases knowledge of their collision-generated entanglement from the overall many-body state.    
Such erasure thus acts to inherently post-select quantum coherent molecules in the remaining sample, a process we dub as \textit{loss-induced quantum autoselection}. 
The result is a suppression of the contrast decay that takes effect on time scales long enough that molecules can undergo more than one collision, therefore motivating the use of the stretched exponential in Eq.~(\ref{eq:stretched_exponential}).

We see this dynamical autoselection effect manifest in 
Fig.~\ref{fig:contrast_loss_vs_noloss}, 
when comparing the contrast evolution from Monte Carlo simulations in a sample with nonzero singlet loss ($\eta_s = 0.05$), to that with zero singlet loss ($\eta_s = 0$).
This representative comparison assumes a gas of $N = 400$ molecules at initial temperature $T = 300$ nK, with an applied electric field ${\rm E} = 12.7$ kV/cm. 
A fit of Eq.~(\ref{eq:stretched_exponential}) to the data in Fig.~\ref{fig:contrast_loss_vs_noloss} gives a stretching parameter of $\zeta = 1.06$ for the case with no loss, but $\zeta = 0.669$ for the case with loss.
Being conditioned on an initial entangling collision, these subsequent lossy $s$-wave collisions are necessarily a multi-body process.

We remark that the barriered and unbarriered nature of triplet and singlet interactions at the level of single partial waves could act as a probe of quantum statistics and entanglement in studies of ultracold chemistry \cite{Liu22_AR, Devolder25_arxiv}, where the affectation of chemical products by quantum state preparation of reactants have already been experimentally observed \cite{Liu24_Sci}.   
Viewed through the lens of quantum information processing, 
singlet loss acts as a form of two-qubit quantum erasure channel \cite{Grassl97_PRA}, discarding maximally entangled singlet Bell states but leaving individual states in the triplet sector unchanged up to a phase.

In general, the interplay of loss-induced autoselection of pure states and interaction-induced dephasing could result in complicated density dependences of the contrast decay. 
In fact, the experimental observation of negative $\kappa$ values indicates the presence of an additional density dependent dephasing mechanisms currently not modeled, that causes the Ramsey contrast decay to decrease with increasing density (\ref{eq:linear_contrastDecayDensityTrend}). 
A potential culprit might be that the experiment actually occurs in a stacked geometry of several quasi-2D pancake layers spaced by $a_{L} = 540$ nm, instead of just a single layer in isolation. 
This interlayer spacing is much smaller than the intralayer mean free path of $\approx 9$ $\mu$m at the largest electric field considered by JILA-KRb. 
Therefore, interlayer long-range dipolar interactions between molecules that are coincident in the $x,y$-plane can be significant, with such occurrences increasing at higher densities. Although these coincident-interlayer trajectories could lead to decoherence, these molecules bear no risk of short-range loss as they remain vertically separated, so that intralayer collisions might be more lossy than with an isolated layer.
Under this hypothesis, we would expect that $\kappa$ should be strictly positive in single 2D layer experiments. We provide an estimate of this effect in App.~\ref{app:interlayer_phase_estimation}, although our findings are currently inconclusive, with more experimental measurements and accurate short-range collisional loss models required. 

Indications of genuine many-body effects just discussed lead us to the considerations of the proceeding section of this paper.

\section{ Prospects for coherent spin mixing and random circuit dynamics \label{sec:unitary_dynamics} }

Looking beyond the investigations of JILA-KRb just described, we now express the broader utility of similar itinerant molecular platforms for explorations of quantum coherent many-body spin mixing physics. 
In particular, we show how such systems give native implementations of pseudorandom unitary circuits with all-to-all connectivity and U(1) charge conservation.

\subsection{ All-channel confinement-induced shielding \label{sec:shielding} }

\begin{figure}[ht]
    \centering
    \includegraphics[width=\columnwidth]{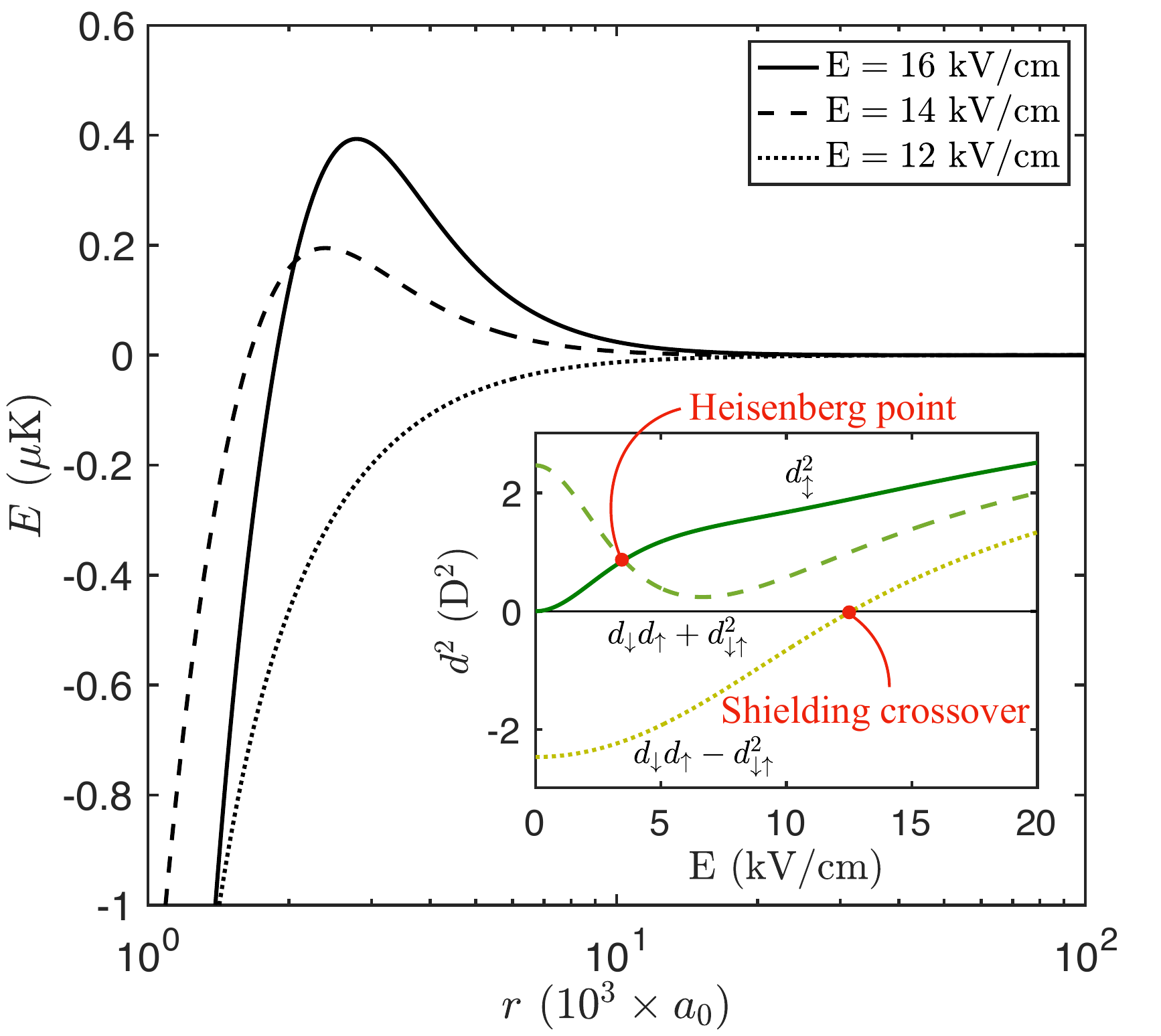}
    \caption{ Lowest adiabatic potential energy curves for singlet state $s$-wave interactions between NaK molecules, for electric fields of ${\cal E} = 12$ kV/cm (dotted), ${\cal E} = 14$ kV/cm (dashed) and ${\cal E} = 16$ kV/cm (solid). The inset plots the squared effective dipole moments for direct, triplet exchange and singlet exchange interaction character as a function of electric field.   }
    \label{fig:NaK_singletPotentials}
\end{figure}

Essential to studying coherent many-body spin dynamics is the suppression of two-body loss to maintain the molecular population and many-body quantum coherence. 
Consequently, we look to other heteronuclear bialkali molecules which have larger dipole moments. For instance, $^{23}$Na$^{40}$K has a body-frame dipole moment of $d = 2.72$ D \cite{Schindewolf22_Nat}, much larger than that of KRb.  
Maintaining a vertical trap with $\omega_z = 2\pi \times 20$ kHz and electric fields of $\gtrsim 12.6$ kV/cm, both achievable in current experiments, the squared singlet dipole moment $d_{\Psi^{-}}^2$ becomes positive in NaK (see the inset of Fig.~\ref{fig:NaK_singletPotentials}), creating a repulsive dipolar barrier to suppress $s$-wave losses. 
The onset of $s$-wave collisional shielding is shown in Fig.~\ref{fig:NaK_singletPotentials}, where by increasing the electric field from ${\rm E} = 12$ to $16$ kV/cm across the zero-crossing of $d_{\downarrow}d_{\uparrow} - d_{\downarrow\uparrow}^2$, a shielding barrier emerges.
The antisymmetry of the singlet state causes molecules to collide with even partial waves, likened to the case of indistinguishable bosons, implying a monotonic increase in the shielding barrier height with larger $\omega_z$ and ${\rm E}$ \cite{Quemener11_PRA}.

\begin{figure}[ht]
    \centering
    \includegraphics[width=\columnwidth]{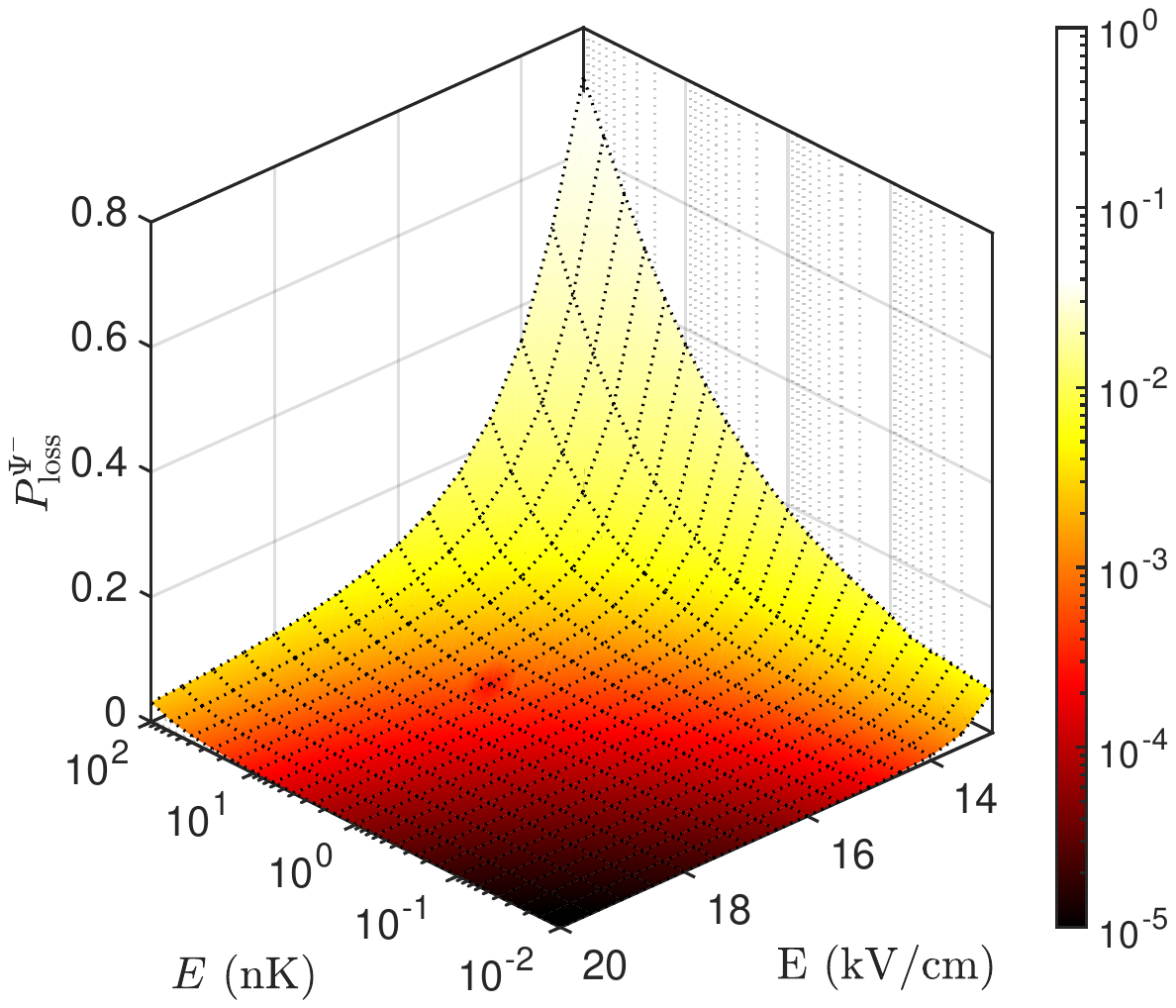}
    \caption{ Singlet-channel loss probability as a function of electric field ${\rm E}$ and collision energy $E$, for quasi-2D confined $^{23}$Na$^{40}$K molecules with $\omega_z = 2\pi \times 20$ kHz. }
    \label{fig:Ploss_vs_Efield_Energy}
\end{figure}

Engineering such a potential barrier should lead to a suppression of molecular loss, which we confirm by computing the expected two-body loss probability from an $s$-wave collision in the singlet channel:
\begin{align}
    P_{\rm loss}^{\Psi^{-}}
    &=
    1 
    -
    | S_{\Psi^{-}, \Psi^{-}}^{0} |^2, 
\end{align}
taken as quench loss from a universal absorbing boundary behind the shielding barrier \cite{Wang15_NJP}.  
We plot $P_{\rm loss}^{\Psi^{-}}$ as a function of electric field and collision energy in Fig.~\ref{fig:Ploss_vs_Efield_Energy}, showing the expected trend of decreasing loss probability with increasing field and decreasing energy. 
We find that for ${\rm E} \geq 16$ kV/cm and $E \lesssim 100$ nK, the loss probability is typically below $10 \%$. 
Written in adimensional terms, we conjecture that all spin polarized $^{1}\Sigma$ bialkali molecules with rotationally dominated van der Waals coefficients \cite{Lepers13_PRA}, can achieve all-channel confinement-induced shielding at $d_0 {\rm E} / B_{\rm rot} > 6$ in sufficiently 2D confined geometries (refer back to Sec.~\ref{sec:single_molecule} for definitions). Notably, KRb has an electronically dominated van der Waals coefficient and requires electric fields of $\gtrsim 30$ kV/cm at $\omega_z = 2\pi \times 100$ kHz confinement for collisional shielding of singlet states.

\subsection{ Unitary circuit representation }

\begin{figure}[ht]
    \centering
    \includegraphics[width=\columnwidth]{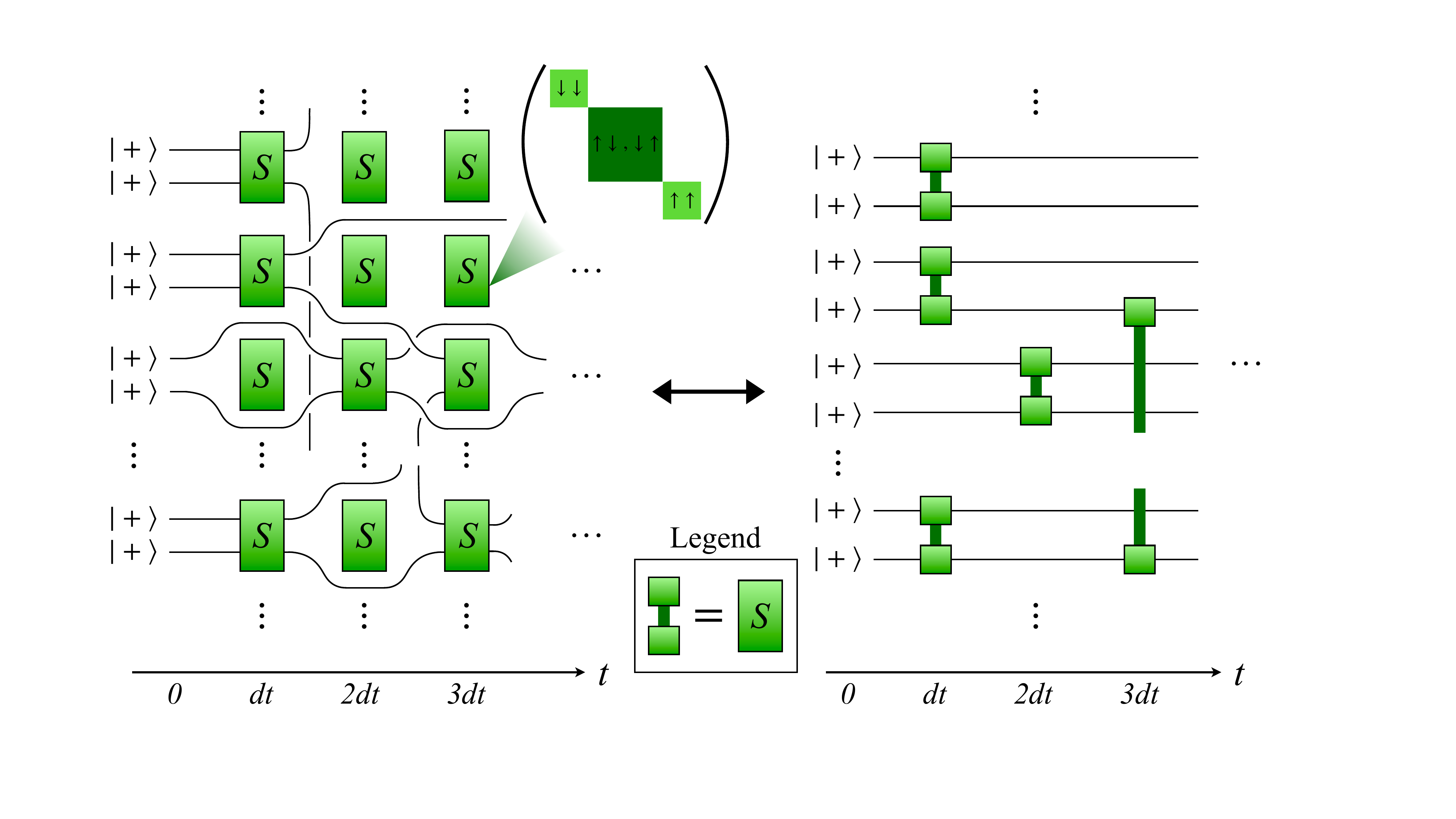}
    \caption{ Equivalent quantum circuit representations of itinerant collisional spin dynamics. The left figure illustrates the MM representation, whereas the right figure illustrates the ATAG representation. Each collisional gate is shown to have a U(1) conserving structure. }
    \label{fig:circuit_picture}
\end{figure}

In this section, we connect the itinerant collisional spin dynamics to a quantum circuit model, usually thought of as idealized models of time evolution on a quantum computer.  
Cast in terms of classical trajectories interspersed with instantaneous collisions, the otherwise continuous spin dynamics is naturally discretized through collision instances, and can be intuitively described in discrete time of step size $dt$, via two equivalent gate-based unitary circuit constructions: (1) ``Mobile molecules" (MM), where the gate architecture is treated as stationary while molecules wander through and between them to experience a cascade of two-molecule gates; (2) ``All-to-all gates" (ATAG), where molecules are treated as stationary while a series of gates are applied with possible all-to-all connectivity.  
In both descriptions, illustrated in Fig.~\ref{fig:circuit_picture}, noise is inherent from the irregularities in time and space of the applied gates.

We emphasize that albeit the MM representation is reminiscent of optical tweezer implementations of quantum circuits \cite{Bluvstein22_Nat, Maskara25_NatPhys}, the motion of spins and gate applications in itinerant molecular systems are not predetermined by user-input, but rather conditional on the quantum state via Eq.~(\ref{eq:scattering_distribution}). 
If external molecular motion is in the strongly ergodic regime as a result of collisional mixing, then the times at which two-molecule gates are applied (i.e. scattering events) will be truly uncorrelated and random. 
We envision that such temporal noise could be controlled by taking the external molecular motion out of equilibrium, as for instance is routinely performed in studies of cross dimensional rethermalization by rapid changes in the ODT \cite{Monroe93_PRL, Goldwin05_PRA, Tang15_PRA, Patscheider22_PRA, Li21_NatPhys}. 
Moreover, although the gates maintain a U(1) conserving structure (\ref{eq:2body_gate_matrix}), their nonzero matrix elements also possess a degree of noise with the collisional phases being dependent on incoming and outgoing scattering momenta (\ref{eq:Vint_Born_approximation}).

On the other hand, the degree of spatial randomness will depend on the diluteness of the sample, where a dilute (collisionally thin) gas would allow molecules to roam mostly unimpeded across the sample width before encountering a collision, permitting highly nonlocal connectivity when collisions do occur.     
Conversely, a hydrodynamic (collisionally thick) sample would greatly restrict free molecular motion and diffusion across the sample width, resulting in more local gate connectivity.  
For our model to apply in a hydrodynamic regime, it will be important to maintain the hierarachy of length scales: $a_d \ll r_{\rm mfp} \ll w_{\rm th}$, where $r_{\rm mfp} = ( n_{\rm 2D} \sigma )^{-1}$ is the molecular mean-free path and $w_{\rm th} = \sqrt{ k_B {\rm T} / ( m \omega_{\perp}^2 ) }$ is the thermal width of the cloud. 
This hierarchy preserves the collisional description of spin dynamics while maintaining a large collision rate compared to the rate at which molecules traverse the trap. 
The ratio ${\rm Kn} = r_{\rm mfp} / w_{\rm th}$ is referred to in the hydrodynamics literature as the Knudsen number \cite{Huang87_JW, Wang23_PRA}, where continuum mechanics applies for ${\rm Kn} \ll 1$, while a kinetic description is more appropriate at ${\rm Kn} \gtrsim 1$. 
Engineering the density by long-wavelength variations in the optical trap surface \cite{Masi21_PRL} could thus serve as a tuning knob for spatial connectivity in the quantum circuit.

To highlight the achievable randomness and all-to-all connectivity, we formulate a continuous time approximation of the circuit dynamics and derive an effective nonlinear Schr\"odinger equation.  
Starting from the many-body spin state $| \Psi_t \rangle = | \Psi(\{ \boldsymbol{\xi}_i \}, t) \rangle$, that
depends parametrically on the phase space coordinates $\boldsymbol{\xi} = (\boldsymbol{q}, \boldsymbol{p})$, its change after a single collision between molecules $i$ and $j$ is given by
\begin{align}
    \ket{ \Psi_{t'} }
    &=
    S^{(i,j)} \ket{ \Psi_t } 
    =
    \left( 
    \mathbb{I} - i T^{(i,j)} 
    \right) 
    \ket{ \Psi_t },
\end{align}
where $S^{(i,j)}$ is the $S$-matrix acting on molecules $i$ and $j$, while $T = i (S - \mathbb{I})$ is the momentum dependent transition $T$-matrix (see App.~\ref{app:multichannel_scattering2D}).  
However, the collisional change in the wavefunction within a time interval $\delta t = t' - t$,  only occurs with probability 
$n_{\rm 2D}( \boldsymbol{q}_i ) 
u( \boldsymbol{q}_i - \boldsymbol{q}_j, \boldsymbol{k}_{i,j} ) 
\beta^{(i, j)}(\boldsymbol{k}_{i,j}) \delta t$, 
based on the expected collision rate constant:
\begin{subequations}
\begin{align}
    \beta^{(i,j)}( \boldsymbol{k}_{i,j} )
    &=
    \sum_{\nu_{i j}}
    \abs{ \bra{ \nu_{i j} }\ket{ \Psi_t } }^2 
    \beta^{\rm el}_{\nu_{i j}}(\boldsymbol{k}_{i,j}), \\
    \beta^{\rm el}_{\nu_{i j}}(\boldsymbol{k}_{i,j}) 
    &=
    \frac{ 2 \pi \hbar }{ m_r }
    \big| 1 - e^{ 2 i \delta_{\nu_{ij}}(\boldsymbol{k}_{i,j})} \big|^2,
\end{align}
\end{subequations}
and the collision coincidence condition: 
\begin{align}
    u(r_{i j}, \boldsymbol{k}_{i,j})
    &=
    \bigg\llbracket
    r_{i j}
    \leq
    \sum_{\nu_{i j}}
    \abs{ \bra{ \nu_{i j} }\ket{ \Psi_t } }^2 
    \sigma_{\nu_{i j}}(\boldsymbol{k}_{i,j})
    \bigg\rrbracket,
\end{align}
where $\boldsymbol{r}_{i j} = \boldsymbol{q}_i - \boldsymbol{q}_j$ and $\hbar \boldsymbol{k}_{i,j} = \boldsymbol{p}_i - \boldsymbol{p}_j$ are the relative coordinate and momentum respectively, while $\llbracket \ldots \rrbracket$ denotes the Iverson bracket. 
Then identifying the variation of the wavefunction from time $t$ to $t'$ by collisions as $\delta \ket{ \Psi_t } = \ket{ \Psi_{t'} } - \ket{ \Psi_t }$, 
the rate of change of the many-body spin state is approximately: 
\begin{align} \label{eq:nonlinear_SchrodingerEquations}
    i  
    \frac{ \delta \ket{ \Psi_t } }{ \delta t }
    &\approx 
    \Bigg[
    \sum_{\langle i,j \rangle} 
    n_{\rm 2D}( \boldsymbol{q}_i ) 
    u\left( 
    \boldsymbol{q}_i - \boldsymbol{q}_j 
    \right) \nonumber\\
    &\quad\quad\:\: \times
    \beta^{(i,j)}( \boldsymbol{k}_{i,j} )
    T^{(i,j)}(\boldsymbol{k}_{i,j})
    \Bigg]
    \ket{ \Psi_t }, 
\end{align} 
identifying an effective Hamiltonian
\begin{subequations} \label{eq:effective_Brownian_Hamiltonian}
\begin{align} 
    {\cal H}(t) 
    &=
    \hbar 
    \sum_{\langle i,j \rangle} 
    \sum_{ \alpha_i, \alpha_j = 0 }^{3}
    \left( 
    \sigma_{\alpha_i}^{(i)} \otimes \sigma_{\alpha_j}^{(j)} 
    \right)
    B_{\alpha_i, \alpha_j}^{(i,j)}(t),  
    \\
    B_{\alpha_i, \alpha_j}^{(i,j)}(t)
    &=
    n_{\rm 2D}( \boldsymbol{q}_i )
    u\left( 
    \boldsymbol{q}_i - \boldsymbol{q}_j, \boldsymbol{k}_{i,j} 
    \right)
    \tau_{i j}(\boldsymbol{k}_{i,j}), \label{eq:Brownian_coefficients}
\end{align}
\end{subequations}
where $\tau_{i j}$ are expansion coefficients of $\beta^{(i,j)}( \boldsymbol{k} ) T^{(i,j)}(\boldsymbol{k})$ into two-site Pauli operators with $\sigma_{0}^{(i)} = \mathbb{I}^{(i)}$. 

The form of $B_{\alpha_i, \alpha_j}^{(i,j)}(t)$ above elucidates that spin dynamics is generated by the $T$-matrix with a characteristic energy scale set by the collision rate. 
In the regime of ergodic molecular motion where collision instances are uncorrelated in time, the structure of ${\cal H}(t)$ in Eq.~(\ref{eq:effective_Brownian_Hamiltonian}) is reminiscent of a Brownian quantum circuit \cite{Lashkari13_JHEP, Zhou19_PRE, Xu19_PRX, Agarwal23_JHEP}: continuous time models where spin-spin couplings fluctuate like Gaussian white noise.  
These models are known, in the absence of symmetries, to be an example of systems that scramble \footnote{ Scrambling refers to a unitary process that maps initial pure product states to macroscopically entangled states. } quantum information in a time logarithmic in the number of DoF, the so-called ``fast scramblers" \cite{Yasuhiro08_JHEP} relating to the black hole complementarity conjecture \cite{tHooft85_NPB, Susskind93_PRD, Brown13_arxiv}. 

Although the U(1) symmetry of the current collisional gates (\ref{eq:2body_gate_matrix}) prevent the coefficients of Eq.~(\ref{eq:Brownian_coefficients}) to be genuine white noise distributed, future studies could consider further engineering of the molecular interactions with microwave dressing \cite{Micheli07_PRA, Gorshkov08_PRL, Lassabliere18_PRL, Karman22_PRA, Quemener23_PRL, Karman25_arxiv}, toward broader explorations of random circuit dynamics.  
As not to distract from the current results in this paper, we leave further numerical analysis of the actual circuit dynamics in these molecular system to a more focused future study. 
Nevertheless, dynamical (as opposed to quenched) disorder induced by finite temperature molecular motion in the continuum connects collisionally stable nondegenerate molecular gases to random quantum circuits, accessible on current experimental platforms.

\section{ Outlook and conclusions \label{sec:conclusion} }

We have formulated a theory of itinerant collisional spin dynamics in ultracold molecular gases. 
At nondegenerate temperatures, the in-plane molecular motion is in the continuum and acts to transport pseudospins around the sample.  
Our model couples the internal (spin) and external (motion) molecular degrees of freedom through scattering events, treating ultracold molecular collisions as the mediators of both spin-spin and spin-motion interactions.    
In performing numerical Monte Carlo simulations, the Ramsey contrast dynamics produced by our model shows quantitative agreement with that measured by the recent JILA KRb experiment, providing both validation to our model and insight to the underlying physics.
Through this analysis, we uncovered the spin-motion coupled phenomenon of loss-induced quantum autoselection, where quantum entanglement can serve to control bimolecular chemistry. 

We then predict that with sufficiently large electric fields, strongly dipolar bialkali molecules can achieve collisional stability in all scattering channels, presenting a pathway towards stable molecular gases for spin mixing experiments.
In a nondegenerate sample, we showed that the itinerant collision spin dynamics can be mapped to a Brownian quantum circuit, expanding the ultracold atomic toolbox for studying quantum chaos and scrambling \cite{Yao16_arxiv, Bentsen19_PRL, Hashizume21_PRL} with ultracold molecules.
Similar spin mixing experiments might also be achieved with magnetic atoms, where tight confinement into quasi-2D layers have been shown to result in significant suppression of spin relaxation \cite{Barral24_NatCom}.

As is relevant to this work of bulk gases, we point to recent experiments that have achieved in-situ spatial correlation measurements of atoms in the continuum \cite{Xiang25_PRL, Yao25_PRL}. By ``freezing" the atomic distribution through quenching on an optical lattice, these experiments are able to take snapshots of site-resolved atomic positions with a quantum gas microscope \cite{Bakr09_Nat, Rosenberg22_NatPhys}.     
We envision that with a similar ``freezing" protocol, site-resolved snapshots of the many-body spin state \cite{Christakis23_Nat} could also be taken in the continuum, possibly allowing for tomography of the entangled state \cite{Aaronson18_ACM, Huang20_NatPhys, Kokail21_NatPhys, Hu23_PRR}.

Finally, our results suggest several avenues for further theoretical and experimental exploration. 
For one, the generation of many-body entanglement could be used to study its effects on state-to-state chemistry. A possible scheme for such a probe could be to first freely evolve the molecules for some time in the all-channel shielded regime. Then an electric field quench to the Heisenberg point could ``pause" dynamics in the now decoherence free triplet sector, whilst lowering the singlet shielding     barrier to promote short-range sticking or reactive dynamics. 
With advancements in quantum state resolved measurements of chemical products \cite{Liu24_Sci}, the distribution over individual quantum states of these products resultant from the initial many-body entangled state could be measured. 

Pertaining to quantum circuit studies, increasing the gas density could have experiments explore the role of hydrodynamic excitations \cite{Wang22_PRA, Wang23_PRA0, Wang23_PRA, Wang25_PRA} on the transport of quantum information in the spin-sector across the gas. With density affecting both the degree of local disorder and circuit connectivity, it will be important to characterize its role in operator spreading \cite{Khemani18_PRX, Nahum18_PRX, Schuster23_PRL} and the information scrambling rates achievable in itinerant molecular platforms. Comparisons can also be made to molecules confined to quasi one-dimensional geometries with 2D optical lattices \cite{Tang18_PRX, Carroll25_Sci}, permitting only nearest neighbor collisional connectivity.       
If cooled to quantum degeneracy, such itinerant dipolar systems could also be utilized to explore spin transport dynamics \cite{Peter12_PRL, Koschorreck13_NatPhys}, unconventional superfluid pairing \cite{Bruun08_PRL, Lee17_PRA}, and the generation of metrologically useful spin-squeezed states \cite{Bilitewski21_PRL, Wellnitz24_PRR}.

\begin{acknowledgments}

This material is based upon work supported by the Nation Science Foundation through a grant for ITAMP. 
J. L. Bohn acknowledges funding from the JILA Physics Frontier Center, Grant PHY-2317149.
The authors would like to thank A. N. Carroll, J. Lin, D. Wellnitz, J. Ye, J. L. Beckey, F. Machado, C. Kokail, P. J. D. Crowley and B. Xing for their invaluable insights.       
The authors especially acknowledge the JILA KRb team for inspiring discussions, and for sharing their experimental data. 

\end{acknowledgments}

\appendix


\section{ Scattering formalism in two dimensions \label{app:multichannel_scattering2D} }

This appendix section is concerned with the definitions and corresponding conventions for the scattering amplitude, matrices and cross sections in two dimensions (2D). 
We first address the differential cross section $d\sigma / d\phi$, in which $d\sigma$ represents the ratio of the number of particles scattered into 2D volume $r dr d\phi$ per unit time, over the number of incident particles into a 2D scattering volume segment per unit cross sectional length segment per unit time.
So for the scattering ansatz \cite{Sadeghpour00_JPB}:
\begin{align} \label{eq:multichannel_scattering_ansatz}
    \ket{ \psi_{\nu \rightarrow \nu'}( \boldsymbol{r} ) }
    &\underset{ k r \gg 1 }{ \rightarrow }
    e^{ i \boldsymbol{k}_{\nu} \cdot \boldsymbol{r} } 
    \ket{ \nu }
    +
    \frac{ e^{ i k_{\nu'} r } }{ \sqrt{ r } }
    f^{ \nu, \nu' }( \boldsymbol{k}_{\nu}, \phi )
    \ket{ \nu' },
\end{align}
where $\boldsymbol{k}_{\nu} = \hat{\boldsymbol{k}}_{\nu} \sqrt{ 2 m_r ( E + \epsilon_{\nu} ) } / \hbar$ is the wave-vector incident with respect to the scattering channel $\nu$ with threshold $\epsilon_{\nu}$, 
the denominator is simply computed as the incident flux of particles:
\begin{align}
    \abs{ \boldsymbol{F}_{\rm in} }
    &=
    \frac{ \hbar }{ m_r }
    \abs{
    {\rm Im}\left\{ \psi_{\rm in}^* \grad \psi_{\rm in} \right\}
    } \nonumber\\
    &=
    \frac{ \hbar }{ m_r }
    \abs{
    {\rm Im}\left\{ e^{ -i \boldsymbol{k}_{\nu} \cdot \boldsymbol{r} } \grad e^{ i \boldsymbol{k}_{\nu} \cdot \boldsymbol{r} } \right\} 
    }
    =
    \frac{ \hbar k_{\nu} }{ m_r }. 
\end{align}
As for the numerator, the differential number of particles scattered into a 2D volume $r d r d\phi$ is given by the probability density multiplied by the differential volume:
\begin{align}
    d N 
    &=
    \abs{ \frac{ e^{ i k_{\nu'} r } }{ \sqrt{ r } }
    f^{ \nu, \nu' }( \boldsymbol{k}_{\nu}, \phi ) }^2 r dr d\phi \nonumber\\
    &=
    \abs{ f^{ \nu, \nu' }( \boldsymbol{k}_{\nu}, \phi ) }^2 dr d\phi,
\end{align}
which taken per unit time gives
\begin{align}
    \frac{ d N }{ d t }
    &=
    v_r 
    \frac{ d N }{ d r } \nonumber\\ 
    &=
    \frac{ \hbar k_{\nu'} }{ m_r }
    \frac{ d N }{ d r } \nonumber\\ 
    &=
    \frac{ \hbar k_{\nu'} }{ m_r } 
    \abs{ f^{ \nu, \nu' }( \boldsymbol{k}_{\nu}, \phi ) }^2 d\phi.
\end{align}
Therefore, the differential cross section for scattering into channel $\nu'$ from channel $\nu$, works out to be
\begin{align} \label{eq:DCS_2D}
    d \sigma^{ \nu \rightarrow \nu' }
    &=
    \frac{ d N / d t }{ \abs{ \boldsymbol{F}_{\rm in} } } \nonumber\\
    &=
    \left( \frac{ \hbar k_{\nu} }{ m_r } \right)^{-1}
    \left( \frac{ \hbar k_{\nu'} }{ m_r } 
    \abs{ f^{ \nu, \nu' }( \boldsymbol{k}_{\nu}, \phi ) }^2 d\phi \right) \nonumber\\
    &=
    \frac{ k_{\nu'} }{ k_{\nu} }
    \abs{ f^{ \nu, \nu' }( \boldsymbol{k}_{\nu}, \phi ) }^2 
    d\phi.
\end{align}

Now to compute the scattering amplitude,  
we first consider an energy eigensolution of 2 free molecules in relative coordinates, the planewave. In 2D, the planewave is expanded into its various partial wave components as
\begin{align}
    e^{ i \boldsymbol{k} \cdot \boldsymbol{r} }
    &=
    2 \pi
    \sum_{ m = -\infty }^{ \infty }
    i^m
    \frac{ e^{ -i m \phi_k } }{ \sqrt{ 2 \pi } }
    J_m (k r)
    \frac{ e^{ i m \phi } }{ \sqrt{ 2 \pi } }.
\end{align}
For lighter notation, we denote partial waves in this appendix section with quantum number $m$ instead of $m_{\phi}$ as in the main text, where there is no ambiguity with molecular mass since it does not appear in this section.  
In the asymptotic limit of $k r \gg 1$ can, up to an overall normalization which plays no role in obtaining the relevant collision quantities, be written as
\begin{widetext}
\begin{align} \label{eq:planewave_expansion}
    e^{ i \boldsymbol{k} \cdot \boldsymbol{r} }
    &\underset{ k r \gg 1 }{ \rightarrow }
    2 \pi
    \sum_{ m = -\infty }^{ \infty }
    i^m
    \frac{ e^{ -i m \phi_k } }{ \sqrt{ 2 \pi } }
    \sqrt{ \frac{ 2 }{ \pi k r } } \cos( k r - \frac{ \pi }{ 4 } ( 2 m + 1 ) )
    \frac{ e^{ i m \phi } }{ \sqrt{ 2 \pi } } \nonumber\\
    &\quad = 
    2 \pi
    \sum_{ m = -\infty }^{ \infty }
    i^m
    \frac{ e^{ -i m \phi_k } }{ \sqrt{ 2 \pi } }
    \left( 
    \frac{ e^{ -i ( k r - \pi ( 2 m + 1 ) / 4 ) } }{ \sqrt{ 2 \pi k r } }
    +
    \frac{ e^{ i ( k r - \pi ( 2 m + 1 ) / 4 ) }  }{ \sqrt{ 2 \pi k r } }
    \right)
    \frac{ e^{ i m \phi } }{ \sqrt{ 2 \pi } },
\end{align}
identifying ingoing $e^{ -i ( k r - \pi ( 2 m + 1 ) / 4 ) }$ and outgoing $e^{ i ( k r - \pi ( 2 m + 1 ) / 4 ) }$ circular wave components. 
We define the $\ket{ m }$ states with a factor $1/\sqrt{ 2 \pi }$ so that they are unit normalized $\bra{ m' }\ket{ m } = \delta_{m', m}$.
If the two initially free molecules now encounter a single-channel scattering event in which they come close, experience a non-negligible interaction potential, then fly off to infinity once again, the wavefunction $\psi_0(\boldsymbol{r})$ will inevitably be modified into
\begin{align} \label{eq:asymptotic_solution}
    \psi( \boldsymbol{r} )
    &\underset{ k r \gg 1 }{ \rightarrow }
    {
    2\pi
    \sum_{ m }
    i^m
    \frac{ e^{ -i m \phi_k } }{ \sqrt{ 2 \pi } }
    \frac{ e^{ -i ( k r - \pi ( 2 m + 1 ) / 4 ) } }{ \sqrt{ 2 \pi k r } }
    \frac{ e^{ i m \phi } }{ \sqrt{ 2 \pi } }
    } 
    \quad\quad\quad\quad
    \quad
    \Bigg\}\:\: \psi_{\rm in}( \boldsymbol{r} )
    \nonumber\\
    &\quad\quad
    +
    {
    2 \pi 
    \sum_{ m, m' }
    i^{m}
    \frac{ e^{ -i m \phi_k } }{ \sqrt{ 2 \pi } }
    S_{m, m'}
    \frac{ e^{ i ( k r - \pi ( 2 m' + 1 ) / 4 ) }  
    }{ \sqrt{ 2 \pi k r } }
    \frac{ e^{ i m' \phi } }{ \sqrt{ 2 \pi } }
    }, 
    \quad\Bigg\}\:\: { \psi_{\rm out}( \boldsymbol{r} ) },
\end{align}
where the ingoing circular waves are free particle solutions and therefore left unscathed, but the outgoing circular waves are modified by a unitary scattering ($S$) matrix, $S_{m, m'}$.
Extending the wavefunction above to the multichannel case, a collision that takes the molecules not only from partial wave $m \rightarrow m'$, but also channel $\nu \rightarrow \nu'$, is represented by the asymptotic wavefunction solution
\begin{align} \label{eq:multichannel_Smatrix_wavefunction}
    \ket{ \psi^{\nu \rightarrow \nu'}_{m \rightarrow m'}( \boldsymbol{r} ) }
    &\underset{ k r \gg 1 }{ \rightarrow }
    \sqrt{ \frac{ 2 \pi }{ k_{\nu'} r } }
    i^{m} 
    \frac{ e^{ -i m \phi_{k} } }{ \sqrt{ 2 \pi } }
    \bigg[
    \delta_{ m, m' } 
    \delta_{\nu, \nu'}
    e^{ -i ( k_{\nu} r - \pi ( 2 m + 1 ) / 4 ) } 
    +
    S_{m, m'}^{\nu, \nu'}
    e^{ i ( k_{\nu'} r - \pi ( 2 m' + 1 ) / 4 ) }
    \bigg]
    \frac{ e^{ i m' \phi } }{ \sqrt{ 2 \pi } }
    \ket{ \nu' }.
\end{align}
As written, $\phi_k$ retains its use as the incident collision orientation, while $\phi$ is now the scattering angle. 
This is the definition of $S_{m, m'}^{\nu, \nu'}$ we shall proceed to work with, from which all other scattering quantities will be defined. 
Subtracting Eq.~(\ref{eq:planewave_expansion}) with $\boldsymbol{k}_{\nu}$ from Eq.~(\ref{eq:multichannel_Smatrix_wavefunction})
then gives
\begin{align}
    & 
    \ket{ \psi^{\nu \rightarrow \nu'}_{m \rightarrow m'}( \boldsymbol{r} ) }
    -
    2 \pi i^m
    \frac{ e^{ -i m \phi_k } }{ \sqrt{ 2 \pi } }
    \sqrt{ \frac{ 2 }{ \pi k_{\nu} r } } \cos( k_{\nu} r - \frac{ \pi }{ 4 } ( 2 m + 1 ) )
    \frac{ e^{ i m \phi } }{ \sqrt{ 2 \pi } } 
    \ket{ \nu } \nonumber\\
    =& \: 
    \sqrt{ \frac{ 2 \pi  }{ k_{\nu'} r } }
    i^{m} 
    \frac{ e^{ -i m \phi_{k} } }{ \sqrt{ 2 \pi } }
    \bigg[
    \delta_{m, m'} 
    \delta_{\nu, \nu'}
    e^{ -i ( k_{\nu} r - \pi ( 2 m + 1 ) / 4 ) }
    +
    S_{m, m'}^{\nu, \nu'}
    e^{ i ( k_{\nu'} r - \pi ( 2 m' + 1 ) / 4 ) }
    \bigg]
    \frac{ e^{ i m' \phi } }{ \sqrt{ 2 \pi } }
    \ket{ \nu' } \nonumber\\
    &\quad -
    \sqrt{ \frac{ 2 \pi }{ k_{\nu'} r } }
    i^{m} 
    \frac{ e^{ -i m \phi_{k} } }{ \sqrt{ 2 \pi } }
    \left[ 
    \delta_{m, m'}
    \delta_{\nu, \nu'}
    \left(
    e^{ -i ( k_{\nu'} r - \pi ( 2 m + 1 ) / 4 ) }
    +
    e^{ i ( k_{\nu'} r - \pi ( 2 m' + 1 ) / 4 ) }
    \right)
    \right]
    \frac{ e^{ i m' \phi } }{ \sqrt{ 2 \pi } }
    \ket{ \nu' } \nonumber\\
    =& \:
    \sqrt{ \frac{ 2 \pi }{ k_{\nu'} r } }
    i^{m} 
    \frac{ e^{ -i m \phi_{k} } }{ \sqrt{ 2 \pi } }
    \left[
    \left(
    S_{m, m'}^{\nu, \nu'}
    -
    \delta_{m, m'}
    \delta_{\nu, \nu'}
    \right)
    e^{ i ( k_{\nu'} r - \pi ( 2 m' + 1 ) / 4 ) }
    \right]
    \frac{ e^{ i m' \phi } }{ \sqrt{ 2 \pi } }
    \ket{ \nu' },
\end{align}
which if compared to the scattering ansatz of Eq.~(\ref{eq:multichannel_scattering_ansatz}) and expanded in the $\ket{ m }$ basis
\begin{align} 
    \ket{ \psi_{\nu \rightarrow \nu'}( \boldsymbol{r} ) }
    &\underset{ k r \gg 1 }{ \rightarrow }
    e^{ i \boldsymbol{k}_{\nu} \cdot \boldsymbol{r} } 
    \ket{ \nu }
    +
    \left(
    \sum_{m, m'}
    \frac{ e^{ -i m' \phi_{k'} } }{ \sqrt{ 2 \pi } }
    f^{ \nu, \nu' }_{m, m'}
    \frac{ e^{ i m \phi_{k} } }{ \sqrt{ 2 \pi } }
    \right)
    \frac{ e^{ i k_{\nu'} r } }{ \sqrt{ r } }
    \ket{ \nu' },
\end{align}
\end{widetext}
identifies the scattering amplitude as
\begin{align} \label{eq:scatamp_vs_Tmatrix}
    f^{ \nu, \nu' }_{m, m'} 
    &=
    e^{ - i \pi / 4 }
    \sqrt{ \frac{ 2 \pi }{ k_{\nu'} } }
    \left[
    S_{m, m'}^{\nu, \nu'}
    -
    \delta_{m, m'}
    \delta_{\nu, \nu'}
    \right]
    \nonumber\\
    &=
    -e^{ i \pi / 4 } 
    \sqrt{ \frac{ 2 \pi }{ k_{\nu'} } }
    T_{m, m'}^{\nu, \nu'} ,
\end{align}
where 
\begin{align}
    T_{m, m'}^{\nu, \nu'} 
    =
    i
    \left[
    S_{m, m'}^{\nu, \nu'} 
    -
    \delta_{m, m'}
    \delta_{\nu, \nu'}
    \right].
\end{align}
The differential cross section for the transition $\nu \rightarrow \nu'$ is then given as
\begin{align}
    \frac{ d \sigma^{\nu \rightarrow \nu'} }{ d \phi_{k'} }
    &=
    \frac{ 2 \pi }{ k_{\nu} }
    \bigg|
    \sum_{m, m'}
    \frac{ e^{ -i m' \phi_{k'} } }{ \sqrt{ 2 \pi } }
    T_{m, m'}^{\nu, \nu'} 
    \frac{ e^{ i m \phi } }{ \sqrt{ 2 \pi } }
    \bigg|^2,
\end{align}
while the total cross section is
\begin{align}
    \sigma^{\nu \rightarrow \nu'}(\phi)
    &=
    \int d\phi_{k'}
    \frac{ d \sigma^{\nu \rightarrow \nu'} }{ d \phi_{k'} } \\
    &=
    \frac{ 2 \pi }{ k_{\nu} }
    \sum_{\tilde{m}, \tilde{m}', m}
    \frac{ e^{ -i \tilde{m} \phi } }{ \sqrt{ 2 \pi } }
    \left( T_{\tilde{m}, \tilde{m}'}^{\nu, \nu'}  \right)^*
    T_{m, \tilde{m}'}^{\nu, \nu'} 
    \frac{ e^{ i m \phi } }{ \sqrt{ 2 \pi } }. \nonumber
\end{align}
and the integral cross section is
\begin{align}
    \overline{\sigma}^{\nu \rightarrow \nu'}
    &=
    \int \frac{ d\phi }{ 2 \pi }
    \int d\phi_{k'}
    \frac{ d \sigma^{\nu \rightarrow \nu'} }{ d \phi_{k'} } \nonumber\\
    &=
    \frac{ 1 }{ k_{\nu} }
    \sum_{m, m'}
    \abs{ T_{m, m'}^{\nu, \nu'}  }^2.
\end{align}
Therefore, we find that the integral elastic, inelastic and quenching cross sections are given as
\begin{subequations}
\begin{align}
    \overline{\sigma}^{\rm el}_{\nu}
    &=
    \frac{ 1 }{ k_{\nu} }
    \sum_{m, m'}
    \abs{ \delta_{m, m'} - S_{m, m'}^{\nu, \nu}  }^2, \\
    \overline{\sigma}^{\rm inel}_{\nu \rightarrow \nu'}
    &=
    \frac{ 1 }{ k_{\nu} }
    \sum_{\nu' \neq \nu}
    \sum_{m, m'}
    \abs{ S_{m, m'}^{\nu, \nu'}  }^2, \\
    \overline{\sigma}^{\rm qu}_{\nu}
    &=
    \frac{ 1 }{ k_{\nu} }
    \sum_{m, m'}
    \left(
    \delta_{m, m'}
    -
    \abs{ S_{m, m'}^{\nu, \nu}  }^2
    \right),
\end{align}
\end{subequations}
respectively. 

If, instead, we were to write Eq.~(\ref{eq:multichannel_Smatrix_wavefunction}) in terms of standing wave solutions:
\begin{widetext}
\begin{align} \label{eq:multichannel_Kmatrix_wavefunction}
    \ket{ \psi^{\nu \rightarrow \nu'}_{m \rightarrow m'}( r ) }
    &\underset{ k r \gg 1 }{ \rightarrow }
    \sqrt{ \frac{ 2 \pi }{ k_{\nu'} r } }
    i^{m} 
    \frac{ e^{ -i m \phi_{k} } }{ \sqrt{ 2 \pi } }
    \bigg[
    \delta_{ m, m' } 
    \delta_{\nu, \nu'}
    \cos( k_{\nu} r - \frac{ \pi ( 2 m + 1 ) }{ 4 } ) 
    +
    K_{m, m'}^{\nu, \nu'}
    \sin( k_{\nu} r - \frac{ \pi ( 2 m + 1 ) }{ 4 } )
    \bigg]
    \frac{ e^{ i m' \phi } }{ \sqrt{ 2 \pi } }
    \ket{ \nu' },
\end{align}
we identify a reaction matrix $K_{m, m'}^{\nu, \nu'}$, which via a expansion into complex exponentials:
\begin{align} 
    \ket{ \psi^{\nu \rightarrow \nu'}_{m \rightarrow m'}( r ) }
    &\underset{ k r \gg 1 }{ \rightarrow }
    \sqrt{ \frac{ 2 \pi }{ k_{\nu'} r } }
    i^{m} 
    \frac{ e^{ -i m \phi_{k} } }{ \sqrt{ 2 \pi } }
    \bigg[
    \frac{1}{2}
    \left(
    \delta_{ m, m' } \delta_{\nu, \nu'}
    +
    i K_{m, m'}^{\nu, \nu'}
    \right) e^{ - i ( k_{\nu'} r - \pi ( 2 m' + 1 ) / 4 ) }
    \nonumber\\
    &\quad\quad\quad\quad\quad\quad\quad\quad\quad\quad +
    \frac{1}{2}
    \left(
    \delta_{ m, m' } \delta_{\nu, \nu'}
    -
    i K_{m, m'}^{\nu, \nu'}
    \right)
    e^{ i ( k_{\nu'} r - \pi ( 2 m' + 1 ) / 4 ) }
    \bigg]
    \frac{ e^{ i m' \phi } }{ \sqrt{ 2 \pi } }
    \ket{ \nu' },
\end{align}
\end{widetext}
and comparison with Eq.~(\ref{eq:multichannel_Smatrix_wavefunction}), identifies the $S$ and $K$-matrix relation
\begin{align}
    \boldsymbol{S}
    &=
    \frac{ \boldsymbol{K}
    -
    i
    \boldsymbol{I} }{ \boldsymbol{K}
    + 
    i
    \boldsymbol{I}}.
\end{align}
The $K$-matrix is what we directly obtain from numerical scattering calculations.


\subsection{ Numerical scattering solutions \label{app:numerical_scattering} }

To obtain the $K$-matrix, we numerically solve the radial Schr\"odinger equation of (\ref{eq:radial_SE}). 
In this work, we utilize a quasi-adiabatic coupling scheme
that utilizes the quasi-adiabatic eigenstates $\Phi(z; r)$, satisfying:
\begin{align} 
    \left(
    -\frac{\hbar^2}{2 m_r}
    \frac{ \partial^2 }{ \partial z^2 }
    +
    V_{\nu, \nu}(r,z)
    +
    V_{{\rm ext}, z}(z)
    \right)
    \Phi_{n_z}(z; r) & \nonumber\\
    =
    W_{n_z}(r)
    \Phi_{n_z}(z; r) &,
\end{align}
where $W_{n_z}(r)$ are the quasi-adiabatic eigenenergies that parametrically depend on $r$ and $\Phi_{n_z}(z; r)$ are the adiabatic channel functions labeled by an axial quantization index $n_z$. 
Above, we have already implicitly assumed specific $\nu$ and $m_{\phi}$ quantum numbers, so we will drop these labels on states and matrix elements for the remainder of this section. 
Utilizing a discrete variable representation (DVR) along $z$, the quasi-adiabatic eigenstates and energies inherit a quantization index $n_z$,
so that Eq.~(\ref{eq:radial_SE}) becomes \cite{Child74_AP}
\begin{align} 
    \delta_{n'_z, n_z} 
    & \left(
    \frac{ \partial^2 }{ \partial r^2 }
    +
    {\cal K}_{n_z}^2(r)
    -
    \frac{ m_{\phi}^2 - 1/4 }{ r^2 }
    \right)
    u(r) \nonumber\\
    &\quad\quad =
    \left[
    Q_{n'_z, n_z}(r)
    \frac{ \partial }{ \partial r }
    +
    R_{n'_z, n_z}(r)
    \right]
    u(r), 
\end{align}
where 
\begin{subequations}
\begin{align}
    Q_{n'_z, n_z}(r)
    &=
    -2 \int d z
    \left[
    \Phi_{n'_z}(z; r)
    \frac{ \partial \Phi_{n_z}(z; r) }{ \partial r }
    \right], \\
    R_{n'_z, n_z}(r)
    &=
    -\int d z
    \left[
    \Phi_{n'_z}(z; r)
    \frac{ \partial^2 \Phi_{n_z}(z; r) }{ \partial r^2 }
    \right], \\
    {\cal K}_{n_z}^2(r)
    &=
    { 2 m_r [ E - \epsilon_{\nu} - W_{n_z}(r) ] / \hbar^2 }.
\end{align}
\end{subequations}
With the collisions in consideration being barriered with collision energies far below the height of the barrier, the wavefunction will have little support in the region of $r$ where $Q_{n'_z, n_z}(r)$ and $R_{n'_z, n_z}(r)$ are appreciable. We will, therefore, ignore these terms which reduces the scattering equations to just a single-channel one:
\begin{align} \label{eq:single_channel_radialSE}
    \left(
    \frac{ \partial^2 }{ \partial r^2 }
    +
    {\cal K}_{n_z}^2(r)
    -
    \frac{ m_{\phi}^2 - 1/4 }{ r^2 }
    \right)
    u(r)
    &\approx 
    0. 
\end{align}
For asymptotically large values of $r$, the quasi-adiabatic eigenenergies tend to $W_{n_z}(r) \rightarrow \hbar \omega_{\perp} n_z$ (ignoring the zero point energy), while the eigenstates tend to quantum harmonic oscillator solutions along $z$.
For this paper, we will only consider the lowest adiabat $W_{0}(r)$ and offset its threshold to zero.  

To perform numerical propagation of Eq.~(\ref{eq:single_channel_radialSE}), we define the quantities
\begin{subequations}
\begin{align}
    {\cal D}_{n_z}
    &=
    \frac{ 1 }{ 2 }
    \frac{ \partial^2 }{ \partial {r}^2 }, \\
    {\cal W}_{n_z}
    &=
    \left(
    \frac{ 1/4 - m_{\phi}^2 }{ 2 {r}^2 }
    +
    \frac{ {k}^2 }{ 2 }
    \right)
    -
    \frac{ m_r }{ \hbar^2 }
    W_{n_z}(r),
\end{align}
\end{subequations}
so that Eq.~(\ref{eq:single_channel_radialSE}) can be recast as the compact system of equations:
\begin{align} \label{eq:concise_radial_equation}
    \left[
    {\cal D}_{n_z}(r)
    +
    {\cal W}_{n_z}(r)
    \right]
    u(r)
    =
    0,
\end{align}
or in terms of the log derivative ${\cal Z}(r) = \chi'(r) \chi^{-1}(r)$:
\begin{align}
    {\cal Z}'(r)
    +
    {\cal Z}^2(r)
    +
    2 {\cal W}_{n_z}(r)
    =
    0.
\end{align}
We opt to obtain solutions to Eq.~(\ref{eq:concise_radial_equation}) by numerical propagation, which requires us to match the solutions at large $\rho$ to the asymptotic solutions of
\begin{align}
    \lim_{r \rightarrow \infty}
    \left(
    \frac{ 1 }{ 2 }
    \frac{ \partial^2 }{ \partial {r}^2 }
    +
    \frac{ 1/4 - m_{\phi}^2 }{ 2 {r}^2 }
    +
    \frac{ {k}^2 }{ 2 }
    \right)
    u(r)
    =
    0,
\end{align}
computed in energy normalized form as
\begin{subequations}
\begin{align}
    f_{m_{\phi}}({r})
    &=
    \sqrt{ \frac{ 2 m_r }{ \hbar^2 {k} } }
    \sqrt{ {k} r } \:
    J_{m_{\phi}}( {k} {r} ), \\
    g_{m_{\phi}}({r})
    &=
    \sqrt{ \frac{ 2 m_r }{ \hbar^2 {k} } }
    \sqrt{ {k} r } \:
    Y_{m_{\phi}}( {k} {r} ),
\end{align}
\end{subequations}
where $J_{m_{\phi}}(z)$ and $Y_{m_{\phi}}(z)$ are Bessel functions of the first and second kind respectively. 
We then propagate Eq.~(\ref{eq:concise_radial_equation}) with an adaptive step size version of the Johnson log-derivative propagation method \cite{Johnson73_JCP}, where we start the propagation and match to the asymptotic solutions at 
\begin{subequations}
\begin{align}
    r_{\rm start}
    &=
    a_{\rm vdW}, \\
    r_{\rm match}
    &=
    10 
    \left(
    \frac{ 2 }{ k_0^2 }
    \right)^{1/3}
    +
    50
    \left(
    \frac{ \max\{ m_{\phi} \}^2 - 1/4 }{ k_0^2 }
    \right)^{1/2},
\end{align}
\end{subequations}
respectively, where $\max\{ m_{\phi} \}$ is the largest partial wave used in calculation. 
The latter condition has its first term set by matching the collision energy to the dipole-dipole interaction energy, while the second term is 50 times the radius at which the collision energy intersects the angular momentum barrier with $\max\{ m_{\phi} \}$.

\section{ Collision partner statistics in JILA-KRb \label{app:collision_stats} }

\begin{figure}[ht]
    \centering
    \includegraphics[width=\columnwidth]{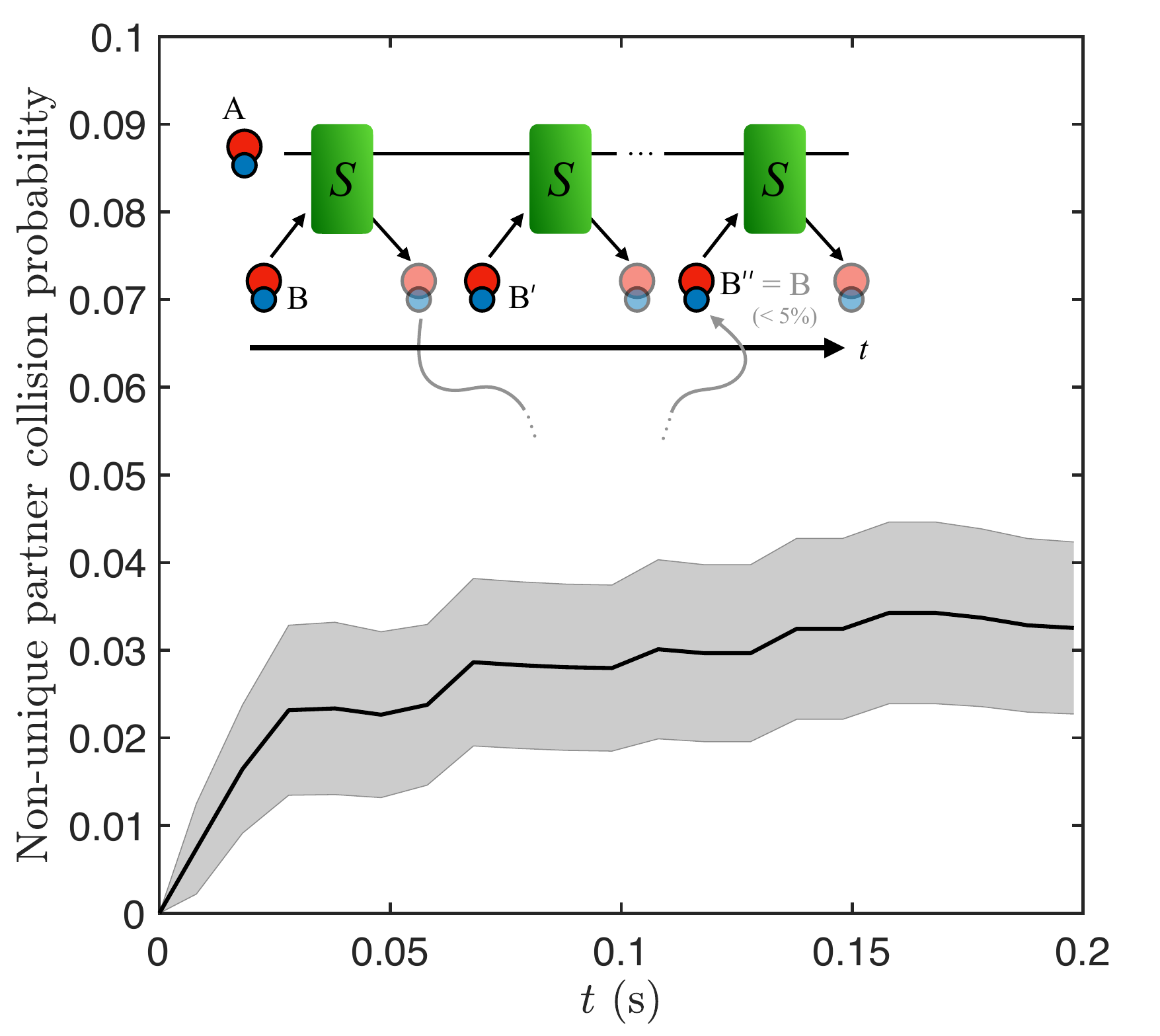}
    \caption{ Probability that a collision partner was previously encountered as a function of time. The inset gives an illustration of the approximately Markov process undergone by molecule A (upper circuit wire), where entangling gates (\ref{eq:2body_gate_matrix}) are applied between collision partners over time, for which collision partners are unique with probability $> 0.95$. }
    \label{fig:Punique_vs_time}
\end{figure}

To ensure the validity of the Markov approximation in Sec.~\ref{sec:Markov_simulations}, we utilize our Monte Carlo simulation to determine the statistics of non-unique collision-partner encounters. 
By tagging each molecule in the simulation and keeping a record of their collision-partners, we can tabulate the number of repeated collision partners they encounter over time. An exemplary plot of the probability that a collision partner was previously encountered as a function of time, is given in Fig.~\ref{fig:Punique_vs_time} for a gas of $N=500$ KRb molecules at $T = 300$ nK subject to an ${\rm E} = 12.72$ kV/cm electric field.    
From our simulations, we find that the collision partners are $> 95 \%$ unique within the time interval over which experimental measurements are taken.

\section{ Estimation of dynamical phases from interlayer dipolar interactions \label{app:interlayer_phase_estimation} }

In this section, we provide the derivation of a mean field estimate for the molecular decoherence from interlayer dipole-dipole interactions.

With molecules between layers distinguishable by their vertical positions, and notationally differentiated by primed or unprimed coordinates in this appendix section, each interlayer molecular pair will have their product basis spin state incur a time varying dynamical phase
\begin{align}
    e^{i \theta_{\alpha, \alpha'}(t)}
    \approx 
    e^{ -\frac{i}{\hbar} 
    \bra{ \alpha, \alpha' }  
    \int_{0}^{t} V_{\rm dd}( \boldsymbol{r}(\tau) - \boldsymbol{r}'(\tau), a_{L} ) d\tau
    \ket{ \alpha, \alpha' } },
\end{align}
that is not echoed away by dynamical decoupling.
For simplicitly of our estimate, we will ignore all entanglement generated between molecules of different layers, amounting to a mean-field approximation.  
In an ergodic sample, a time average amounts to an ensemble average, so we can approximate the dynamical phase accumulated by any one molecule as
\begin{align} \label{eq:interlayer_phase}
    \theta_{\alpha, \alpha'}(t)
    &\approx
    -\frac{ t }{ \hbar }
    \int \frac{ d^2\boldsymbol{r} d^2\boldsymbol{r}' }{ 2 N_{\rm  mol} }
    n_{\rm 2D}(\boldsymbol{r})
    n'_{\rm 2D}(\boldsymbol{r}') \nonumber\\
    &\quad\quad\quad\quad \times 
    \bra{ \alpha, \alpha' }  
    V_{\rm dd}( \boldsymbol{r} - \boldsymbol{r}', a_{L} )
    \ket{ \alpha, \alpha' }
    \nonumber\\
    &\approx 
    \frac{ N'_{\rm mol} t }{ 32 \pi \hbar }
    \Theta(\vartheta)
    \frac{ d_{ \alpha} d_{ \alpha' } }{ 4 \pi \epsilon_0 a_{L}^3 } , \nonumber
\end{align}
where $\vartheta = k_B T / ( m \omega_{\perp}^2 a_{L}^2 )$ compares the thermal energy to a natural interlayer trapping energy scale, and $\Theta(\vartheta) = 
    \sqrt{ \pi \vartheta }
    \left[ 
    { ( 2 \vartheta + 1 ) / \vartheta^3 }
    \right] 
    {\rm Erfc}\left( \frac{ 1 }{ 2 \sqrt{ \vartheta } } \right)
    e^{ \frac{ 1 }{ 4 \vartheta } }
    -
    ( 2 / \vartheta^2 )$ is a geometric function. 
Evident from Eq.~(\ref{eq:interlayer_phase}), $\theta_{\alpha, \alpha'}(t)$ scales linearly with $N'_{\rm mol}$ and therefore increases with density over a fixed time interval.

Accounting for dynamical decoupling, two molecules $A$ and $A'$ in separated (primed and unprimed) layers that undergo grazing trajectories will have their initial two-body state $\ket{+}\otimes\ket{+}$, evolve into $[e^{i \vartheta_{\downarrow\downarrow}} ( \ket{ \downarrow\downarrow } + \ket{ \uparrow\uparrow } ) + e^{i \vartheta_{\downarrow\uparrow}} ( \ket{ \downarrow\uparrow } + \ket{ \uparrow\downarrow } )] / 2$.
The same process occurs for all other molecules, so that the singlet channel component between molecule $A$ and another intralayer molecule $B$ in state $\varrho_B(t) \approx \varrho_A(t)$ is:  
\begin{align}
    \bra{ \Psi^{-} } \varrho_A(t) \otimes \varrho_B(t) \ket{ \Psi^{-} }
    &\approx 
    \frac{ 1 }{ 4 } 
    \sin^2[\vartheta_{\downarrow\downarrow}(t) - \vartheta_{\downarrow\uparrow}(t)], 
\end{align}
while the contrast of each molecule has changed by $\Delta C(t) \approx 1 - \cos[\vartheta_{\downarrow\downarrow}(t) - \vartheta_{\downarrow\uparrow}(t)]$.
Adopting the experimental parameters of $N'_{\rm mol} = 400$, $T = 300$ nK, $\omega_{\perp} = 39$ Hz and ${\rm E} = 12.72$ kV/cm,
we get $(\vartheta_{\downarrow\downarrow} - \vartheta_{\downarrow\uparrow})/t \approx 0.25$ s$^{-1}$ which is around 30 times smaller than the intralayer contrast decay rate $\Gamma$ (\ref{eq:contrast_decay_rate}). 

\nocite{*}
\bibliography{main.bib} 

\end{document}